\documentstyle[aps,pra]{revtex}
\begin{document}
\draft

\newcommand{\uu}[1]{\underline{#1}}
\newcommand{\pp}[1]{\phantom{#1}}
\newcommand{\be}{\begin{eqnarray}}
\newcommand{\ee}{\end{eqnarray}}
\newcommand{\ve}{\varepsilon}
\newcommand{\vs}{\varsigma}
\newcommand{\Tr}{{\,\rm Tr\,}}
\newcommand{\pol}{\frac{1}{2}}

\title{
Non-canonical quantum optics (II):
Poincar\'e covariant formalism and thermodynamic
limit
}
\author{Marek~Czachor and Monika Syty}
\address{
Katedra Fizyki Teoretycznej i Metod Matematycznych\\
Politechnika Gda\'{n}ska,
ul. Narutowicza 11/12, 80-952 Gda\'{n}sk, Poland}
\maketitle

\begin{abstract}
The paper contains further development of the idea of field quantization
introduced in M. Czachor, J. Phys. A: Math. Gen. {\bf 33} (2000) 8081-8103.
The formalism is extended to the relativistic domain. The link to the
standard theory is obtained via a thermodynamic limit.
Unitary representations of the Poincar\'e group at the level of fields and
states are explicitly given. Non-canonical multi-photon and coherent
states are introduced. In the thermodynamic limit the statistics of
photons in a coherent state is Poissonian. The $S$ matrix of radiation
fields produced by a classical current is
given by a non-canonical coherent-state displacement operator, a fact
automatically eliminating the infrared catastrophe. Field operators
are shown to be operators and not operator-valued
distributions, and can be multiplied at the same point in
configuration space. An exactly solvable example is used to compare
predictions of the standard theory with those of non-canonical
quantum optics, and explicitly shows the mechanism of automatic
ultraviolet regularization occuring in the non-canonical theory.
Similar conclusions are obtained in perturbation theory, where one
finds the standard Feynman diagrams, but the Feynman rules are
modified.
A comparison with the Dicke-Hepp-Lieb model allows to identify the
physical structure behind the non-canonical algebra as corresponding
to an ensemble of indefinite-frequency oscillators with constant
density $N/V$.
\end{abstract}
\pacs{PACS: 11.10.-z, 04.60.Ds, 98.80.Es}


\section{Introduction}

The idea
of ``non-canonical quantization" of electromagnetic fields was introduced in
\cite{I}. One begins with the observation that even in
nonrelativistic quantum mechanics it is natural to treat the
frequency $\omega$ characterizing a harmonic oscillator as an
eigenvalue and not a parameter. A replacement of $\omega$ by an
operator $\hat\omega$ leads to an indefinite-frequency oscillator
with altered (non-)canonical commutation
relations, and forms a natural departure point for a new version of
the `old fashioned' field quantization.
An analysis of physical structures
associated with such indefinite-frequency operators was discussed in
detail in \cite{II}. In the present paper we show that the
non-canonically quantized electromagnetic field may be regarded as an
ensemble of indefinite-frequency oscillators of constant density.

We extend the formalism to the relativistic
domain. Representations of both
the non-canonical commutation relations and the Poincar\'e group are
explicitly constructed.
Multi-photon and coherent states are defined. Their properties make
them similar to those from the canonical formalism if one performs an
appropriately defined
thermodynamic limit $N\to\infty$. Vacuum states transform as a
massless scalar field.
Radiation fields produced by a classical current lead to the correct form of
the $S$ matrix. The $S$ matrix is proportional to a
non-canonical coherent-state displacement operator, a fact
eliminating the infrared catastrophe.

Commutation relations for vector potentials are
found showing certain deviations from locality due to nontrivial
structure of vacua but simultaneously allow for multiplication of
field operators at the same point in configuration space, a first
hint suggesting the ultraviolet finiteness of the theory. Two other
arguments in favor of the ultraviolet finiteness are based on
analysis of perturbation theory and the spontaneous emission survival
amplitude.

An example of the spontaneous-emission survival amplitude is exactly
solvable in both canonical and non-canonical theories and can be
employed to illustrate the links between the two approaches. Of
particular interest is the analogy between our method of quantization
and the Hepp-Lieb treatment of the Dicke model. We show that that the
Dicke-Hepp-Lieb Hamiltonian has the structure of the same type as our
multi-oscillator non-canonical Hamiltonian. This leads to the
conclusion that the formal
requirement that the RHS of non-CCR commutators satisfies the
resolution of unity is equivalent to the physical requirement
that the electromagnetic field is an ensemble of indefinite-frequency
harmonic oscillators of constant density $N/V$.

The fact that the RHS of commutation relations is modified by the
presence of an element from the center of the algebra makes our
``non-canonical" fields analogous to the so-called generalized free
fields \cite{Greenberg,BLT}. The essential difference between the two
approaches is that, first of all, our fields do interact with charges
and we do
not demand the Poincar\'e invariance of field commutators but employ
the algebra which is only covariant. As a result we loose
Poincar\'e invariance of vacuum and locality. This
leads to analogies with
nonlocal quantum field theories discussed in \cite{Efimov}.
The loss of locality implies also interesting analogies
between our approach and quantum field theory in non-commutative
space \cite{Jackiw,Cai,geometry}.
Still, there are also differences between what we propose and
these approaches. For example, the
formalism of \cite{Efimov} is based on nonlocal distributions and
regularizations of Hamiltonians. Similarly to our approach vacua in
such theories are Poincar\'e non-invariant, a fact used to
circumvent limitations imposed by the Haag theorem
\cite{Efimov}. However, in our approach the nonlocal
distributions occur at the level of amplitudes or averages, and not
at the algebra of field operators. Analogously, perturbative
expansions of amplitudes are automatically regularized in spite of
the fact that there is no regularization at the level of operators.

\section{Notation}

In order to control covariance properties of fields in generalized
frameworks it is best to work in a manifestly covariant formalism.
The most convenient is the one based on spinors and passive unitary
transformations.

\subsection{Spinor convention and fields}

We take $c=1$ and $\hbar=1$.
The index notation we use in the paper is consistent with the
Penrose-Rindler spinor and world-tensor convention \cite{PR}. The
electromagnetic field-tensor and its dual are
\be
F_{ab}
&=&
\left(
\begin{array}{cccc}
0    & E_1 & E_2  & E_3\\
-E_1 & 0   & -B_3 & B_2 \\
-E_2 & B_3 & 0 & -B_1\\
-E_3 & -B_2 & B_1 & 0
\end{array}
\right)\\
^*F_{ab}
&=&
\left(
\begin{array}{cccc}
0    & -B_1 & -B_2  & -B_3\\
B_1 & 0   & -E_3 & E_2 \\
B_2 & E_3 & 0 & -E_1\\
B_3 & -E_2 & E_1 & 0
\end{array}
\right)
\ee
Self-dual and anti-self-dual parts of $F_{ab}$ are related to the
electromagnetic spinor by
\be
^+ F_{ab}
&=&
\frac{1}{2}
\Big(F_{ab}-i ^* F_{ab}\Big)=\varepsilon_{AB}\bar \varphi_{A'B'}
=
\left(
\begin{array}{cccc}
0    & F_1 & F_2  & F_3\\
-F_1 & 0   & iF_3 & -iF_2 \\
-F_2 & -iF_3 & 0 & iF_1\\
-F_3 & iF_2 & -iF_1 & 0
\end{array}
\right)\\
^- F_{ab}
&=&
\frac{1}{2}
\Big(F_{ab}+i ^* F_{ab}\Big)=\varepsilon_{A'B'}\varphi_{AB}
\ee
where $\bbox F=(\bbox E+i\bbox B)/2$ is the Riemann-Silberstein
vector \cite{IBB-pwf}. Denote $k\cdot x=k_ax^a$.
The electromagnetic spinor has the following Fourier representation
\cite{Woodhouse,IZBB}
\be
\varphi_{AB}(x)
&=&
\int d\Gamma(\bbox k)\pi_{A}(\bbox k)\pi_{B}(\bbox k)
\Big(f(\bbox k,-)e^{-ik\cdot x}+\overline{f(\bbox k,+)}e^{ik\cdot
x}\Big)\label{varphi}\\
\bar\varphi_{A'B'}(x)
&=&
\int d\Gamma(\bbox k)\bar\pi_{A'}(\bbox k)\bar\pi_{B'}(\bbox k)
\Big(f(\bbox k,+)e^{-ik\cdot x}
+\overline{f(\bbox k,-)}e^{ik\cdot x}\Big)\label{varphi'}
\ee
where the spinor field $\pi_{A}(\bbox k)$ is related to
the future-pointing 4-momentum by
\be
k^a&=& \pi^{A}(\bbox k)\bar \pi^{A'}(\bbox k)=(k_0,\bbox k)
=(|\bbox k|,\bbox k)
\ee
and the invariant measure on the light-cone is $d\Gamma(\bbox k)=
\big[(2\pi)^3 2k_0\big]^{-1}d^3k$.
Anti-self-dual and self-dual parts of the field tensor are
\be
^- F_{ab}(x)
&=&
\int d\Gamma(\bbox k)\varepsilon_{A'B'}\pi_{A}(\bbox k)\pi_{B}(\bbox k)
\Big(f(\bbox k,-)e^{-ik\cdot x}
+\overline{f(\bbox k,+)}e^{ik\cdot x}\Big)\\
&=&
\int d\Gamma(\bbox k)e_{ab}(\bbox k)
\Big(f(\bbox k,-)e^{-ik\cdot x}
+\overline{f(\bbox k,+)}e^{ik\cdot x}\Big)\\
^+ F_{ab}(x)
&=&
\int d\Gamma(\bbox k)\varepsilon_{AB}\bar\pi_{A'}(\bbox
k)\bar\pi_{B'}(\bbox k)
\Big(\overline{f(\bbox k,-)}e^{ik\cdot x}
+f(\bbox k,+)e^{-ik\cdot x}\Big)\\
&=&
\int d\Gamma(\bbox k)\bar e_{ab}(\bbox k)
\Big(\overline{f(\bbox k,-)}e^{ik\cdot x}
+f(\bbox k,+)e^{-ik\cdot x}\Big)
\ee
The latter form is used by the Bia{\l}ynicki-Birulas in \cite{IZBB}.

The sign in the amplitude $f(\bbox k,\pm)$ corresponds to the value
of helicity of positive-frequency fields.

The four-vector potential $A_a(x)$ is related to the electromagnetic spinor by
\be
\varphi_{XY}(x)
&=&
\nabla{_{(X}}{^{Y'}}A{_{Y)Y'}}(x)\label{A}
\ee
In the Lorenz gauge $\nabla^aA_a=0$ we do not have to symmetrize the unprimed
indices and
\be
\varphi_{XY}(x)
&=&
\nabla{_{X}}{^{Y'}}A{_{YY'}}(x).
\ee
One of the possible Lorenz gauges is
\be
A_a(x)
&=&
i\int d\Gamma(\bbox k)\Bigg(
m_{a}(\bbox k)
\Big(f(\bbox k,+)e^{-ik\cdot x}-\overline{f(\bbox k,-)}e^{ik\cdot x}\Big)
-
\bar m_{a}(\bbox k)
\Big(\overline{f(\bbox k,+)}e^{ik\cdot x}-f(\bbox k,-)e^{-ik\cdot x}\Big)
\Bigg)
\label{gauge}
\ee
where $\omega_A\pi^A=1$, i.e $\omega_A=\omega_A(\bbox k)$ is a
spin-frame partner of $\pi^A(\bbox k)$.
In (\ref{gauge}) we have introduced the null vectors
\be
m_{a}(\bbox k)
&=&
\omega_A(\bbox k)\bar\pi_{A'}(\bbox k)\\
\bar m_{a}(\bbox k)
&=&
\pi_A(\bbox k)\bar\omega_{A'}(\bbox k)
\ee
which, together with
\be
k_a &=&
\pi_A(\bbox k)\bar\pi_{A'}(\bbox k)\\
\omega_a(\bbox k)
&=&
\omega_A(\bbox k)\bar\omega_{A'}(\bbox k)
\ee
form a null tetrad \cite{PR}.

A change of gauge is in the
Fourier domain represented by a shift by
a multiple of $k^a$. The form (\ref{gauge}) shows that gauge freedom
is related to the nonuniqueness of $\omega_A(\bbox k)$ which can be
shited by a multiple of $\pi_A(\bbox k)$.

\subsection{Momentum representation}

Consider the momentum-space basis normalized by
\be
\langle \bbox p|\bbox p'\rangle=(2\pi)^3 2p_0\delta^{(3)}(\bbox
p-\bbox p') =\delta_\Gamma(\bbox p,\bbox p'), \quad p_0>0.
\ee
The identity operator in momentum space is
$\int d\Gamma(\bbox p)|\bbox p\rangle\langle \bbox p|$.
We can use the following explicit realization of
$|\bbox p\rangle=|f_p\rangle$ in terms of distributions
\be
f_p(\bbox k)
&=&
(2\pi)^32 p_0 \delta^{(3)}(\bbox p-\bbox k)
=
\delta_\Gamma(\bbox p,\bbox k)
\ee
Since
\be
\int d\Gamma(\bbox k) F(\bbox k)\delta_\Gamma(\bbox p,\bbox k)
&=&
\int d^3k \delta^{(3)}(\bbox p-\bbox k)F(\bbox k)=F(\bbox p)
\ee
the Fourier transform of $f_p(\bbox k)$ is
\be
\check f_p(x)
&=&
\int d\Gamma(\bbox k)f_p(\bbox k) e^{-i k\cdot x}=
e^{-i p\cdot x}
\ee
If $1$ is the identity operator occuring at the right-hand-side of CCR
$[a_s,a^{\dag}_{s'}]=\delta_{ss'}1$, we denote
\be
I_{\bbox k}
&=&
|\bbox k\rangle\langle \bbox k|\otimes 1,\\
I
&=&
\int d\Gamma(\bbox k)|\bbox k\rangle\langle \bbox k|\otimes 1.
\ee

\subsection{Multi-particle conventions}

Let $A$ be an operator $A: {\cal H}\to {\cal H}$ where $\cal H$ is a
one-particle Hilbert space. The multi-particle Hilbert space
\be
\uu {\cal H}=\oplus_{n=1}^\infty \otimes_s^n {\cal H}
\ee
is the Hilbert space of states corresponding to an indefinite number
of bosonic particles; $\otimes_s^n {\cal H}$ stands for a space of
symmetric states in $\underbrace{{\cal H}\otimes\dots \otimes {\cal H}}_n$.
We introduce the following notation for operators defined at the
multi-particle level:
\be
\oplus_{\alpha_n} A
&=&
\alpha_1 A
\oplus
\alpha_2\big(A\otimes I+I\otimes A\big)\oplus
\alpha_3\big(A\otimes I\otimes I+
I\otimes A \otimes I+
I\otimes I\otimes A\big)\oplus
\dots
\ee
Here $\oplus_{\alpha_n} A: \uu {\cal H}\to \uu {\cal H}$,
$\alpha_n$ are real or complex parameters, and $I$ is the identity
operator in $\cal H$.

The following properties follow directly from the definition
\be
[\oplus_{\alpha_n} A,\oplus_{\beta_n} B]&=&\oplus_{\alpha_n\beta_n}
[A,B]\\
e^{\oplus_{\alpha_n} A}
&=&
\oplus_{n=1}^\infty
\underbrace{e^{\alpha_n A}\otimes\dots \otimes e^{\alpha_n A}}_n\\
e^{\oplus_1 A}\oplus_{\beta_n} Be^{-\oplus_1 A}
&=&
\oplus_{\beta_n}e^{A}Be^{-A}
\ee
Identity operators  in $\uu{\cal H}$ and $\cal H$ are related by
\be
\uu I &=& \oplus_{\frac{1}{n}} I.
\ee
We will often use the operator
\be
\uu I_{\bbox k}
&=&
\oplus_{\frac{1}{n}} I_{\bbox k}.
\ee

\section{Poincar\'e transformations of classical electromagnetic fields}

Denote, respectively, by $\Lambda$ and $y$ the $SL(2,C)$ and
4-translation parts of a Poincar\'e transformation\cite{PT}
$(\Lambda,y)$.
The spinor representation of the Poincar\'e group acts in the space
of anti-self-dual electromagnetic fields in 4-position representation
as follows:
\be
^-\hat F_{ab}(x)
&\mapsto&
\big(T_{\Lambda,y}{^-}\hat F\big)_{ab}(x)\\
&=&
\Lambda{_a}{^c}\Lambda{_b}{^d}{^-}\hat F_{cd}\big(\Lambda^{-1}(x-y)\big)\\
&=&
\int d\Gamma(\bbox k)\varepsilon_{A'B'}\Lambda{_A}{^C}\pi_{C}(\bbox k)
\Lambda{_B}{^D}\pi_{D}(\bbox k)
\Big(f(\bbox k,-)e^{-ik\cdot \Lambda^{-1}(x-y)}
+\overline{f(\bbox k,+)}e^{ik\cdot \Lambda^{-1}(x-y)}\Big)\\
&=&
\int d\Gamma(\bbox k)\varepsilon_{A'B'}\Lambda{_A}{^C}\pi_{C}(\bbox
{\Lambda^{-1}k})
\Lambda{_B}{^D}\pi_{D}(\bbox {\Lambda^{-1}k})
\Big(f(\bbox {\Lambda^{-1}k},-)e^{-ik\cdot (x-y)}
+\overline{f(\bbox {\Lambda^{-1}k},+)}e^{ik\cdot (x-y)}\Big)
\ee
where $\bbox{\Lambda^{-1}k}$ is the spacelike part of
$\Lambda^{-1}{_a}{^b}k{_b}$.
The transformed field
\be
(\Lambda\pi){_{A}}(\bbox k)
=
\Lambda{_A}{^C}\pi_{C}(\bbox {\Lambda^{-1}k})
\ee
satisfies
\be
k^a&=& \pi^{A}(\bbox k)\bar \pi^{A'}(\bbox k)
=
(\Lambda\pi){^{A}}(\bbox k)\overline{\Lambda\pi}{^{A'}}(\bbox k)
\ee
Now, if $\omega_{A}(\bbox k)$ is a spin-frame partner of
$\pi_{A}(\bbox k)$, i.e. $\omega_{A}(\bbox k)\pi^{A}(\bbox k)=1$,
one can write
\be
\pi^{A}(\bbox k)
&=&
(\Lambda\pi){^{A}}(\bbox k)\bar \omega_{A'}(\bbox
k)\overline{\Lambda\pi}{^{A'}}(\bbox k)
\ee
which shows that
$\pi^{A}(\bbox k)$ and $(\Lambda\pi){_{A}}(\bbox k)
=
\Lambda{_A}{^C}\pi_{C}(\bbox {\Lambda^{-1}k})$
are proportional to each other, the proportionality factor being
\be
\lambda(\Lambda,\bbox k)
=\bar \omega_{A'}(\bbox
k)\overline{\Lambda\pi}{^{A'}}(\bbox k).
\ee
The form (\ref{gauge}) showed that the gauge freedom is related to
shifts
\be
\omega_A(\bbox k)\mapsto \omega_A(\bbox k)+{\rm scalar}\times
\pi_A(\bbox k)
\ee
which do not affect $\lambda(\Lambda,\bbox k)$ making it independent
of gauge.
Using again
\be
k^a&=&|\lambda(\Lambda,\bbox k)|^2 \pi^{A}(\bbox k)\bar \pi^{A'}(\bbox k)
=|\lambda(\Lambda,\bbox k)|^2k^a
\ee
one concludes that $\lambda(\Lambda,\bbox k)$ is a phase factor
\be
\lambda(\Lambda,\bbox k)=e^{i\Theta(\Lambda,\bbox k)}
\ee
and we find
\be
\big(T_{\Lambda,y}{^-}\hat F\big)_{ab}(x)
&=&
\int d\Gamma(\bbox k)\varepsilon_{A'B'}\pi_{A}(\bbox {k})
\pi_{B}(\bbox {k})e^{-2i\Theta(\Lambda,\bbox k)}
\Big(f(\bbox {\Lambda^{-1}k},-)e^{-ik\cdot (x-y)}
+\overline{f(\bbox {\Lambda^{-1}k},+)}e^{ik\cdot (x-y)}\Big)\\
&=&
\int d\Gamma(\bbox k)
e_{ab}(\bbox {k})
\Big(e^{-2i\Theta(\Lambda,\bbox k)}e^{ik\cdot y}
f(\bbox {\Lambda^{-1}k},-)e^{-ik\cdot x}
+
e^{-2i\Theta(\Lambda,\bbox k)}e^{-ik\cdot y}
\overline{f(\bbox {\Lambda^{-1}k},+)}e^{ik\cdot x}\Big)\\
&=&
\int d\Gamma(\bbox k)
e_{ab}(\bbox {k})
\Big(
(T_{\Lambda,y}f)(\bbox {k},-)e^{-ik\cdot x}
+
\overline{(T_{\Lambda,y}f)(\bbox {k},+)}e^{ik\cdot x}\Big)
\ee
We have obtained therefore the {\it passive\/} transformation of the
classical wave function
\be
f(\bbox {k},\pm)
\mapsto
(T_{\Lambda,y}f)(\bbox {k},\pm)
=
e^{\pm 2i\Theta(\Lambda,\bbox k)}e^{ik\cdot y}
f(\bbox {\Lambda^{-1}k},\pm).\label{unit}
\ee
which is simply the unitary zero-mass spin-1 representation
the Poincar\'e group. The above derivation clearly shows that the
rule (\ref{unit}) is obtained without any particular assumption about
the choice of $f(\bbox {k},\pm)$. In particular, the derivation
remains valid even if one replaces functions $f$ by operators,
independently of their algebraic properties.
The above `passive' viewpoint on the structure of unitary
representations is particularly useful if one aims at generalizations
of CCR. The passive derivation of all the non-tachyonic unitary
representations of the Poincar\'e group can be found in \cite{MC-BW}.

\section{Non-canonical quantization}

We follow the strategy described in \cite{I} and \cite{II}. Let $a_s$
be {\it canonical\/} annihilation operators satisfying CCR
$[a_s,a^{\dag}_{s'}]=\delta_{ss'}1$.
Define the 1-oscillator {\it non-canonical\/} creation and
annihilation operators \cite{uwaga}
\be
a(f)^{\dag}
&=&
\sum_s\int d\Gamma(\bbox k)f(\bbox k,s)|\bbox k\rangle\langle \bbox
k| \otimes a^{\dag}_s\\
&=&
\sum_s\int d\Gamma(\bbox k)f(\bbox k,s)a(\bbox k,s)^{\dag}\label{a(f)}\\
a(f)
&=&
\sum_s\int d\Gamma(\bbox k)\overline{f(\bbox k,s)}
|\bbox k\rangle\langle \bbox k| \otimes a_s\\
&=&
\sum_s\int d\Gamma(\bbox k)\overline{f(\bbox k,s)}
a(\bbox k,s)\label{a(f)^+}
\ee
satisfying the non-CCR algebra
\be
[a(\bbox k,s),a(\bbox k',s')^{\dag}]
&=&
\delta_{ss'}\delta_\Gamma(\bbox k, \bbox k')
|\bbox k\rangle\langle \bbox k| \otimes 1\\
&=&
\delta_{ss'}\delta_\Gamma(\bbox k, \bbox k')
I_{\bbox k}.
\ee
Taking, in particular, $f_{p,r}(\bbox k,s)=\delta_{rs}
\delta_\Gamma(\bbox p,\bbox k)$ one finds
\be
a(f_{p,r})
&=&
|\bbox p\rangle\langle \bbox p| \otimes a_r
=
a(\bbox p,r).\label{53}
\ee
The one-oscillator quantization is
\be
^-\hat F_{ab}(x)
&=&
\varepsilon_{A'B'}\hat \varphi_{AB}(x)\\
&=&
\int d\Gamma(\bbox k)\varepsilon_{A'B'}\pi_{A}(\bbox k)\pi_{B}(\bbox k)
\Big(a(\bbox k,-)e^{-ik\cdot x}
+a(\bbox k,+)^{\dag}e^{ik\cdot x}\Big)\\
&=&
\int d\Gamma(\bbox k)e_{ab}(\bbox k)
\Big(a(\bbox k,-)e^{-ik\cdot x}
+a(\bbox k,+)^{\dag}e^{ik\cdot x}\Big)
\ee
Spinor transformations of $^-\hat F_{ab}(x)$ lead to the passive
transformation
\be
a(\bbox {k},\pm)
\mapsto
(T_{\Lambda,y}a)(\bbox {k},\pm)
=
e^{\pm 2i\Theta(\Lambda,\bbox k)}e^{ik\cdot y}
a(\bbox {\Lambda^{-1}k},\pm).
\ee
The quantization procedure is gauge independent since we work at the
gauge-independent level of $^-\hat F_{ab}(x)$.

Multi-oscillator fields are defined in terms of
\be
\uu a(\bbox p,s)
&=&
\oplus_\frac{1}{\sqrt{n}}
a(\bbox p,s)\label{58}
\ee
and
\be
\uu a(f)^{\dag}
&=&
\sum_s\int d\Gamma(\bbox k)f(\bbox k,s)\uu a(\bbox k,s)^{\dag}\\
\uu a(f)
&=&
\sum_s\int d\Gamma(\bbox k)\overline{f(\bbox k,s)}
\uu a(\bbox k,s).
\ee
The fact that the coefficients $\frac{1}{\sqrt{n}}$ are found in the
multi-oscillator definition may appear awkward. At least three
different {\it formal\/} arguments for such a choice of the multi-particle
extension of field operators were given in \cite{I}. Below, in Sec.
X, we will show that this very special choice of the non-CCR representation
corresponds {\it physically\/} to an ensemble of oscillators
uniformly distributed in space.

The non-CCR algebra is
\be
[\uu a(f),\uu a(g)^{\dag}]
&=&
\sum_{s}\int d\Gamma(\bbox k)
\overline{f(\bbox k,s)}g(\bbox k,s)
\uu I_{\bbox k}
\ee
The right-hand-side of the above formula is in the center of the
non-CCR algebra, i.e.
\be
\big[[\uu a(f),\uu a(g)^{\dag}],\uu a(h)\big]
&=&
0\\
\big[[\uu a(f),\uu a(g)^{\dag}],\uu a(h)^{\dag}\big]
&=&
0
\ee
Useful is also the formula
\be
[\uu a(f_{p,r}),\uu a(f_{p',r'})^{\dag}]
&=&
[\uu a(\bbox p,r),\uu a(\bbox p',r')^{\dag}]\\
&=&
\delta_{rr'}
\delta_\Gamma(\bbox p,\bbox p')\uu I_{\bbox p}.\label{65}
\ee
The presence of $\uu I_{\bbox p}$ at the right-hand-sides of
non-CCR will influence orthogonality properties of multi-photon
states, as we shall see later.

At the multi-oscillator level the electromagnetic field tensor operator is
\be
{\uu F}_{ab}(x)
&=&
\int d\Gamma(\bbox k)e_{ab}(\bbox k)
\Big(\uu a(\bbox k,-)e^{-ik\cdot x}
+\uu a(\bbox k,+)^{\dag}e^{ik\cdot x}\Big)\\
&+&
\int d\Gamma(\bbox k)\bar e_{ab}(\bbox k)
\Big(\uu a(\bbox k,-)^{\dag}e^{ik\cdot x}
+\uu a(\bbox k,+)e^{-ik\cdot x}\Big).
\ee
The four-potential operator is (in our choice of gauge)
\be
{\uu A}_a(x)
&=&
i\int d\Gamma(\bbox k)\Bigg(
m_{a}(\bbox k)
\Big(\uu a(\bbox k,+)e^{-ik\cdot x}
-\uu a(\bbox k,-)^{\dag}e^{ik\cdot x}\Big)
+
\bar m_{a}(\bbox k)
\Big(\uu a(\bbox k,-)e^{-ik\cdot x}
-\uu a(\bbox k,+)^{\dag}e^{ik\cdot x}\Big)
\Bigg).\label{A_a}
\ee
It is well known that field ``operators" of the standard theory (let
us denote them by ${\hat A}_a(x)$) are
in fact operator-valued distributions. As a consequence the operator
products of the form
$
{\hat A}_a(x){\hat A}_b(y)
$
are ill defined and lead to ultraviolet-divergent expressions if $x=y$.
The techniques of dealing with ultraviolet divergences
are based on appropriate regularizations of products of distributions
taken at ``diagonals" in configuration space \cite{Scharf,Connes}.
In Sec. VI w shall see that the non-canonically quantized ${\uu
A}_a(x)$ is an {\it operator\/} and there is no difficulty with
${\uu A}_a(x){\uu A}_b(x)$.

\section{Action of the Poincar\'e group on field operators}

We are interested in finding the representation of the group in terms
of unitary
similarity transformations, i.e.
\be
\uu a(\bbox {k},\pm)
\mapsto
e^{\pm 2i\Theta(\Lambda,\bbox k)}e^{ik\cdot y}
\uu a(\bbox {\Lambda^{-1}k},\pm)
=
{\uu U}_{\Lambda,y}^{\dag} \uu a(\bbox {k},\pm){\uu U}_{\Lambda,y}
\ee
It is sufficient to find an appropriate representation at the
one-oscillator level. Indeed, assume we have found ${U}_{\Lambda,y}$
satisfying
\be
e^{\pm 2i\Theta(\Lambda,\bbox k)}e^{ik\cdot y}
a(\bbox {\Lambda^{-1}k},\pm)
=
{U}_{\Lambda,y}^{\dag} a(\bbox {k},\pm){U}_{\Lambda,y}.
\ee
Then
\be
{\uu U}_{\Lambda,y}
&=&
\bigoplus_{N=1}^\infty
\underbrace{U_{\Lambda,y}\otimes\dots\otimes U_{\Lambda,y}}_N.
\ee

\subsection{Four-translations}

The definition of four momentum for a single harmonic oscillator is
\be
P_a
&=&
\int d\Gamma(\bbox k)k_a|\bbox k\rangle\langle \bbox k|\otimes h
\ee
where
\be
h=\frac{1}{2}\sum_s\Big(a^{\dag}_sa_s+a_s\,a^{\dag}_s\Big)
=\sum_s h_s.
\ee
One immediately verifies that
\be
e^{iP\cdot x}a(\bbox k,s)e^{-iP\cdot x}
&=&
a(\bbox k,s) e^{-ix\cdot k}\\
e^{iP\cdot x}a(\bbox k,s)^{\dag}e^{-iP\cdot x}
&=&
a(\bbox k,s)^{\dag} e^{ix\cdot k}
\ee
implying
\be
U_{\bbox 1,y}=e^{iy\cdot P}.
\ee
Consequently, the generator of four-translations corresponding to
$
\uu U_{\bbox 1,y}=e^{iy\cdot \uu P}
$
is
$
\uu P_a=\oplus_1 P_a
$
and
\be
e^{i{\uu P}\cdot x}\uu a(\bbox k,s)^{\dag}e^{-i{\uu P}\cdot x}
&=&
\uu a(\bbox k,s)^{\dag} e^{ix\cdot k}\\
e^{i{\uu P}\cdot x}\uu a(\bbox k,s)e^{-i{\uu P}\cdot x}
&=&
\uu a(\bbox k,s) e^{-ix\cdot k}.
\ee
The $x$-dependence of field operators can be introduced via $\uu P$:
\be
{\uu F}_{ab}(x)
&=&
e^{i{\uu P}\cdot x}
{\uu F}_{ab}
e^{-i{\uu P}\cdot x}
\ee
\subsection{Rotations and boosts}

To find an analogous representation of
\be
\uu a(\bbox {k},\pm)
\mapsto
e^{\pm 2i\Theta(\Lambda,\bbox k)}
\uu a(\bbox {\Lambda^{-1}k},\pm)
=
{\uu U}_{\Lambda,0}^{\dag} \uu a(\bbox {k},\pm){\uu U}_{\Lambda,0}
\ee
we define
\be
U_{\Lambda,0}
&=&
\exp\Big(\sum_s 2is\int d\Gamma(\bbox k)\Theta(\Lambda,\bbox k)|\bbox
k\rangle \langle \bbox k|\otimes h_s\Big)
\Big(\sum_r\int d\Gamma(\bbox p)|\bbox p,r\rangle
\langle \bbox {\Lambda^{-1}p},r|\otimes  1\Big).
\ee
Finally the transformations of the field tensor are
\be
{\uu U}_{\Lambda,0}^{\dag}{\uu F}_{ab}(x){\uu U}_{\Lambda,0}
&=&
\Lambda{_a}{^c}\Lambda{_b}{^d}{\uu F}_{cd}(\Lambda^{-1}x)\\
{\uu U}_{\bbox 1,y}^{\dag}{\uu F}_{ab}(x){\uu U}_{\bbox 1,y}
&=&
{\uu F}_{ab}(x-y)
\ee
The zero-energy part of $\uu P$ can be removed by a unitary
transformation leading to a {\it vacuum picture\/} dynamics (cf.
\cite{II}). We will describe this in more detail after having
discussed the properties of non-canonical states.

\section{States and their Poincar\'e transformations}

It is clear that in order to control transformation properties of
states it is sufficient to discuss single-oscillator representations.
We shall start with single-oscillator states and then extend them to
many oscillators.

\subsection{Representation in the one-oscillator sector}

The one-oscillator Hilbert space consists of functions $f$ satisfying
\be
\sum_{n_+,n_-=0}^\infty\int d\Gamma(\bbox k)|f(\bbox k,n_+,n_-)|^2
<\infty.
\ee
We will write them in the Dirac notation as
\be
|f\rangle
&=&
\sum_{n_\pm}\int d\Gamma(\bbox k)f(\bbox k,n_+,n_-)|\bbox k,n_+,n_-\rangle.
\ee
The representation of the Poincar\'e group is
\be
|f\rangle
\mapsto
U_{\Lambda,y}|f\rangle
&=&
U_{\bbox 1,y}U_{\Lambda,0}|f\rangle\nonumber\\
&=&
\sum_{n_\pm}\int d\Gamma(\bbox k)f(\bbox {\Lambda^{-1}k},n_+,n_-)
e^{2i(n_+-n_-)\Theta(\Lambda,\bbox {k})}e^{ik\cdot
y(n_++n_-+1/2)} |\bbox {k},n_+,n_-\rangle.
\ee
The latter formula can be written as
\be
f(\bbox k,n_+,n_-)
\mapsto
U_{\Lambda,y}f(\bbox k,n_+,n_-)
&=&
e^{ik\cdot y(n_++n_-+1/2)}
e^{2i(n_+-n_-)\Theta(\Lambda,\bbox {k})}
f(\bbox {\Lambda^{-1}k},n_+,n_-)\label{U1}
\ee
or
\be
U_{\Lambda,y}|f\rangle
=
|U_{\Lambda,y}f\rangle.\label{U2}
\ee
The form (\ref{U1}) is very similar to the zero-mass spin-1
representation (\ref{unit}), the difference being in the multiplier
$n_++n_-+1/2$. One can check by a straightforward calculation that
(\ref{U1}) defines a representation of the group.

\subsection{Generators and vacuum picture}

Denote by $K_a$ and $L_{ab}+S_{ab}$ the generators of 4-translations
and $SL(2,C)$ of the standard zero-mass spin-1 unitary representation
of the Poincar\'e group. $L_{ab}$ denotes the orbital part of the
generator. The generators of (\ref{U1}) are then
\be
P^a&=&K_a\otimes h\label{gen1}\\
J^{ab}&=&L^{ab}\otimes 1+ S{^{ab}}{_{ss}}\otimes h_s.\label{gen2}
\ee
$S{^{ab}}{_{ss'}}$ are matrix elements of $S{^{ab}}$ (which is a
diagonal $p$-dependent matrix).
Denote by $\tilde S_{ab}$ the generators of the $(1/2,1/2)$ spinor
representation of $SL(2,C)$, i.e.
$\Lambda=\exp\big(i\xi^{ab}\tilde S_{ab}/2\big)$. The generators
of the unitary representation are defined by
\be
P_af(\bbox k,n_+,n_-)
&=&
-i\frac{\partial}{\partial y^a}
U_{\Lambda,y}f(\bbox k,n_+,n_-)\big|_{\xi,y=0}\\
J_{ab}f(\bbox k,n_+,n_-)
&=&
-i\frac{\partial}{\partial \xi^{ab}}
U_{\Lambda,y}f(\bbox k,n_+,n_-)\big|_{\xi,y=0}
\ee
In what follows we will work in a ``vacuum picture", i.e with unitary
transformations
\be
f(\bbox k,n_+,n_-)
\mapsto
V_{\Lambda,y}f(\bbox k,n_+,n_-)
&=&
e^{i(n_++n_-)k\cdot y}
e^{2i(n_+-n_-)\Theta(\Lambda,\bbox {k})}
f(\bbox {\Lambda^{-1}k},n_+,n_-)\label{Uvac}.
\ee
The transition
\be
U_{\Lambda,y}\mapsto V_{\Lambda,y}=
W_{y}^{\dag}U_{\Lambda,y}
\ee
is performed by means of the unitary transformation
which commutes with non-CCR creation and annihilation operators.

Let us stress that the fact that we ``remove" the zero-energy parts
from generators does not mean that energy of vacuum is zero. The
vacuum picture is in a sense a choice of representation co-moving
with vacuum.

\subsection{Vacuum states}

Vacuum states are all the states which are annihilated by all
annihilation operators. At the one-oscillator level these are states
of the form
\be
|O\rangle=\int d\Gamma(\bbox k)O(\bbox k)|\bbox k,0,0\rangle.
\ee
Even in the vacuum picture the vacuum states are not Poincar\'e
invariant since
\be
V_{\Lambda,y}O(\bbox k)
&=&
O(\bbox {\Lambda^{-1}k})
\ee
which means they transform as a 4-translation-invariant
scalar field. We will often meet the expression
$Z(\bbox k)=|O(\bbox k)|^2$ describing the probability density of the
``zero modes".

\subsection{Coherent states}

An analogue of the standard coherent (or ``semiclassical") state is
at the 1-oscillator level
\be
|O_\alpha\rangle
&=&
\int d\Gamma(\bbox k)O(\bbox {k})
|\bbox {k}\rangle|\alpha (\bbox {k},+),\alpha (\bbox {k},-)\rangle
\ee
where
\be
a_s|\alpha (\bbox {k},+),\alpha (\bbox {k},-)\rangle
=
\alpha (\bbox {k},s)|\alpha (\bbox {k},+),\alpha (\bbox {k},-)\rangle
\ee
Its explicit form in the basis of eigenstates of the oscillator is
\be
|O_\alpha\rangle
&=&
\int d\Gamma(\bbox k)
O(\bbox k)\sum_{n_+,n_-=0}^\infty\frac{\alpha(\bbox k,+)^{n_+}}{\sqrt{n_+!}}
e^{-|\alpha(\bbox k,+)|^2/2}
\frac{\alpha(\bbox k,-)^{n_-}}{\sqrt{n_-!}}
e^{-|\alpha(\bbox k,-)|^2/2}
|\bbox k\rangle|n_+\rangle|n_-\rangle\\
&=&
\sum_{n_+,n_-}\int d\Gamma(\bbox k)
O_\alpha(\bbox k,n_+,n_-)
|\bbox k,n_+,n_-\rangle
\ee
where
\be
O_\alpha(\bbox k,n_+,n_-)
=
\frac{1}{\sqrt{n_+!n_-!}}O(\bbox k)
\alpha(\bbox k,+)^{n_+}\alpha(\bbox k,-)^{n_-}
e^{-\sum_\pm|\alpha(\bbox k,\pm)|^2/2}
\ee
The average of the 1-oscillator field operator evaluated in such a
coherent state is
\be
\langle O_\alpha|{^-}\hat F_{ab}(x)|O_\alpha\rangle
&=&
\int d\Gamma(\bbox k)e_{ab}(\bbox k)
Z(\bbox k)
\Big(\alpha(\bbox k,-)e^{-ik\cdot x}
+
\overline{\alpha(\bbox k,+)}e^{ik\cdot x}\Big).
\ee
The Poincar\'e transformation of the state implies
\be
\langle O_\alpha|V_{\Lambda,y}^{\dag}
{^-}\hat F_{ab}(x)V_{\Lambda,y}|O_\alpha\rangle
=
\Lambda{_a}{^c}\Lambda{_b}{^d}
\langle O_\alpha|{^-}\hat F_{cd}\big(\Lambda^{-1}(x-y)\big)|O_\alpha\rangle
\ee
The coherent-state wave function transforms by
\be
V_{\Lambda,y}O_\alpha(\bbox k,n_+,n_-)
&=&
e^{i(n_++n_-)k\cdot y}
e^{2i(n_+-n_-)\Theta(\Lambda,\bbox {k})}
O_\alpha(\bbox {\Lambda^{-1}k},n_+,n_-)\\
&=&
O(\bbox {\Lambda^{-1}k})
\prod_{s=\pm}
e^{in_sk\cdot y}
e^{2isn_s\Theta(\Lambda,\bbox {k})}
\frac{1}{\sqrt{n_s!}}
\alpha(\bbox {\Lambda^{-1}k},s)^{n_s}
e^{-|\alpha(\bbox {\Lambda^{-1}k},s)|^2/2}
\\
&=&
O(\bbox {\Lambda^{-1}k})
\prod_{s=\pm}
\frac{1}{\sqrt{n_s!}}\Big(
T_{\Lambda,y}\alpha(\bbox {k},s)\Big)^{n_s}
e^{-|T_{\Lambda,y}\alpha(\bbox {k},s)|^2/2}
\ee
where
\be
\alpha(\bbox {k},s)\mapsto T_{\Lambda,y}\alpha(\bbox {k},s)
\ee
is the spin-1 massless unitary representation (\ref{unit}).
Using this result we get again
\be
\langle O_\alpha|V_{\Lambda,y}^{\dag}
{^-}\hat F_{ab}(x)V_{\Lambda,y}|O_\alpha\rangle
&=&
\int d\Gamma(\bbox k)e_{ab}(\bbox k)
Z(\bbox {\Lambda^{-1}k})
\Big(T_{\Lambda,y}\alpha(\bbox {k},-)
e^{-ik\cdot x}
+
\overline{T_{\Lambda,y}\alpha(\bbox {k},+)}
e^{ik\cdot x}\Big)\\
&=&
\Lambda{_a}{^c}\Lambda{_b}{^d}
\langle O_\alpha|{^-}\hat F_{cd}\big(\Lambda^{-1}(x-y)\big)|O_\alpha\rangle
\ee
showing that those somewhat counter-intuitive forms of
$U_{\Lambda,y}$ and $V_{\Lambda,y}$ are consistent with passive
$T_{\Lambda,y}$ transformations of classical wave functions.

With $a(\beta)$ and $a(\beta)^{\dag}$ given by
(\ref{a(f)})-(\ref{a(f)^+}) we define the displacement operator
\be
{\cal D}(\beta)
&=&
e^{a(\beta)^{\dag}-a(\beta)}\\
&=&
\exp\Big(\sum_s\int d\Gamma(\bbox k)
\big(\beta(\bbox k,s) a(\bbox k,s)^{\dag}
-\overline{\beta(\bbox k,s)} a(\bbox k,s)\big)\Big)\\
&=&
\int d\Gamma(\bbox k)
|\bbox k,s\rangle\langle\bbox k,s|\otimes
e^{\sum_s\big(\beta(\bbox k,s) a^{\dag}_s
-
\overline{\beta(\bbox k,s)} a_s\big)}
\ee
which performs a shift of the classical wave function
\be
{\cal D}(\beta)|O_\alpha\rangle=|O_{\alpha+\beta}\rangle
\ee
and commutes with $I_{\bbox k}$:
\be
{\cal D}(\beta)^{\dag}I_{\bbox k} {\cal D}(\beta)=I_{\bbox k}.
\ee
Vacuum states are also coherent states corresponding to $\alpha=0$.

\subsection{Multi-oscillator coherent states}

Consider a family $\alpha_N(\bbox {k},s)$, $N=1,2,\dots$ of functions and the
state
\be
|\uu O_\alpha\rangle
&=&
\bigoplus_{N=1}^\infty
\sqrt{p_N}\underbrace{|O_{\alpha_N}\rangle\otimes\dots
\otimes |O_{\alpha_N}\rangle}_N\label{mocs}
\ee
where
\be
|O_{\alpha_N}\rangle
&=&
\int d\Gamma(\bbox k)O(\bbox {k})
|\bbox {k}\rangle|\alpha_N (\bbox {k},+),\alpha_N (\bbox {k},-)\rangle
\ee
and $\sum_{N=1}^\infty p_N=1$.
Taking, for example, $\alpha_N(\bbox {k},s)=\alpha(\bbox {k},s)/\sqrt{N}$
we find
\be
\langle\uu O_\alpha|{^-}{\uu {\hat F}}_{ab}(x)|\uu O_\alpha\rangle
&=&
\int d\Gamma(\bbox k)e_{ab}(\bbox k)
Z(\bbox k)
\Big(\alpha(\bbox k,-)e^{-ik\cdot x}
+
\overline{\alpha(\bbox k,+)}e^{ik\cdot x}\Big)
=
\langle O_\alpha|{^-}\hat F_{ab}(x)|O_\alpha\rangle
\ee
i.e. the same result as in the 1-oscillator case.

A multi-oscillator displacement operator is
\be
{\uu {\cal D}}(\beta)
&=&
\bigoplus_{N=1}^\infty \underbrace{{\cal D}(\beta_N)\otimes
\dots\otimes {\cal D}(\beta_N)}_N
=
e^{\uu a(\beta)^{\dag}-\uu a(\beta)},\label{uu D}
\ee
$\beta_N(\bbox {k},s)=\beta(\bbox {k},s)/\sqrt{N}$,
implying
\be
\uu {\cal D}(\beta)|\uu O_\alpha\rangle
&=&
|\uu O_{\alpha+\beta}\rangle\\
{\uu {\cal D}}(\beta)^{\dag}\uu a(\bbox p,s){\uu {\cal D}}(\beta)
&=&
\uu a(\bbox p,s)+\beta(\bbox p,s)\uu I_{\bbox p}\\
{\uu {\cal D}}(\beta)^{\dag}\uu I_{\bbox p}{\uu {\cal D}}(\beta)
&=&
\uu I_{\bbox p}.
\ee
The fact that $\alpha_N(\bbox {k},s)=\alpha(\bbox {k},s)/\sqrt{N}$
will be shown to be of crucial importance for the question of
statistics of excitations of multi-oscillator coherent states. Let us
note that a similar property of coherent states was found in \cite{I}
when we employed the definition in terms of eigenstates of
annihilation operators.

\subsection{Multi-oscillator vacua}

Vacuum consists of states with $n_\pm=0$, i.e.
with all the oscillators in their ground states.
Of particular interest, due to its simplicity, is the following vacuum state
\be
|\uu O\rangle
&=&
\bigoplus_{N=1}^\infty
\sqrt{p_N}\underbrace{|O\rangle\otimes\dots
\otimes |O\rangle}_N\label{uu O}
\ee
where
\be
|O\rangle
&=&
\int d\Gamma(\bbox k)O(\bbox {k})
|\bbox {k},0,0\rangle
\ee
Such a vacuum is simultaneously a particular case of a coherent state
with $\alpha(\bbox k,s)=0$. Coherent states are related to the vacuum
state via the displacement operator
\be
{\uu {\cal D}}(\alpha)|\uu O\rangle
&=&
\bigoplus_{N=1}^\infty
\sqrt{p_N}\underbrace{|O_{\alpha_N}\rangle\otimes\dots
\otimes |O_{\alpha_N}\rangle}_N\\
&=&
|\uu O_\alpha\rangle.
\ee

\subsection{Normalized 1-photon states}

Consider the vector
\be
\uu a(f)^{\dag}|\uu O\rangle.
\ee
Choosing the particular form (\ref{uu O})
we find
\be
\langle \uu O|\uu a(f)\uu a(g)^{\dag}|\uu O\rangle
&=&
\sum_{s}\int d\Gamma(\bbox k) Z(\bbox k)
\overline{f(\bbox k,s)}g(\bbox k,s)
\\
&=&
\langle fO|gO\rangle=:\langle f|g\rangle_Z.
\ee
$fO$ denotes the pointlike product
$fO(\bbox k,s)=O(\bbox k)f(\bbox k,s)$. Since anyway only the modulus
$|O(\bbox k)|=Z(\bbox k)^{\frac{1}{2}}$
occurs in the above scalar products one can
also work with $f_B(\bbox k,s)=Z(\bbox k)^{\frac{1}{2}}f(\bbox k,s)$.
The relation between $f_B$ and $f$ resembles the one between the bare
and renormalized fields \cite{Weinberg}.
We believe this is more than just an analogy.

Thinking of bases in the Hilbert space
one can take functions $f_i$ satisfying
\be
\langle f_i|f_j\rangle_Z=\delta_{ij}\label{delta}=
\langle f_{Bi}|f_{Bj}\rangle.
\ee
\subsection{Normalization of multi-photon states}

Normalization of multi-photon states is more complicated. In this
section we will discuss this point in detail since the argument we
give is very characteristic for the non-canonical framework. It
will be used in Sec. X to show that in the thermodynamic limit of a
large number of
oscillators the non-CCR perturbation theory tends to the CCR one but
in a version which is automatically regularized. We will
also use a similar trick to show that the multi-oscillator coherent
states have, again in the thermodynamic limit, Poissonian statistics of
excitations.

Denote by $\sum_\sigma$ the sum over all the permutations of the set
$\{1,\dots,m\}$.
\medskip

\noindent
{\bf Theorem 1.} Consider the vacuum state (\ref{uu O}) with $p_N=1$
for some $N$. Then
\be
{}&{}&\lim_{N\to\infty}
\langle \uu O|\uu a(f_1)\dots \uu a(f_m)\uu a(g_1)^{\dag}\dots
\uu a(g_m)^{\dag}|\uu O\rangle
=
\sum_{\sigma}
\langle f_1|g_{\sigma(1)}\rangle_Z
\dots
\langle f_m|g_{\sigma(m)}\rangle_Z\nonumber\\
&{}&\pp{==}=
\sum_{\sigma}\sum_{s_1\dots s_m}\int d\Gamma(\bbox k_1)Z(\bbox k_1)\dots
d\Gamma(\bbox k_m)Z(\bbox k_m)
\overline{f_1(\bbox k_1,s_1)}\dots
\overline{f_m(\bbox k_m,s_m)}
g_{\sigma(1)}(\bbox k_1,s_1)
\dots
g_{\sigma(m)}(\bbox k_m,s_m)\nonumber
\ee
{\it Proof:\/}
The scalar product of two general unnormalized multi-photon states
is
\be
{}&{}&\langle \uu O|\uu a(f_1)\dots \uu a(f_m)\uu a(g_1)^{\dag}\dots
\uu a(g_m)^{\dag}|\uu O\rangle\nonumber\\
&{}&\pp =
=
\sum_{\sigma}\sum_{s_1\dots s_m}\int d\Gamma(\bbox k_1)\dots
d\Gamma(\bbox k_m)\overline{f_1(\bbox k_1,s_1)}\dots
\overline{f_m(\bbox k_m,s_m)}
g_{\sigma(1)}(\bbox k_1,s_1)
\dots
g_{\sigma(m)}(\bbox k_m,s_m)
\langle \uu O|
\uu I_{\bbox k_1}\dots \uu I_{\bbox k_m}
|\uu O\rangle\nonumber\\
&{}&\pp =
=
\sum_{\sigma}\sum_{s_1\dots s_m}\int d\Gamma(\bbox k_1)\dots
d\Gamma(\bbox k_m)\overline{f_1(\bbox k_1,s_1)}\dots
\overline{f_m(\bbox k_m,s_m)}
g_{\sigma(1)}(\bbox k_1,s_1)
\dots
g_{\sigma(m)}(\bbox k_m,s_m)
\nonumber\\
&{}&
\pp =
\times
\frac{1}{N^m}
\underbrace{
\langle O|\dots\langle O|}_N
\Big(I_{\bbox k_1}\otimes \dots\otimes I+
\dots +I\otimes\dots\otimes I_{\bbox k_1}\Big)
\dots
\Big(I_{\bbox k_m}\otimes \dots\otimes I+
\dots +I\otimes\dots\otimes I_{\bbox k_m}\Big)
\underbrace{
|O\rangle\dots |O\rangle}_N\label{multi-norm}
\ee
Further analysis of (\ref{multi-norm}) can be simplified by the
following notation:
\be
1_{k_j} &=&  I_{\bbox k_j}\otimes \dots\otimes I\nonumber
\\
2_{k_j} &=&  I\otimes I_{\bbox k_j}\otimes \dots\otimes I\nonumber
\\
        &\vdots& \nonumber\\
N_{k_j} &=& I\otimes\dots\otimes I_{\bbox k_j}\nonumber
\ee
with $j=1,\dots,m$; the sums-integrals $\sum_{s_j}\int
d\Gamma(\bbox k_j)$ are denoted by $\sum_{k_j}$.  Then
(\ref{multi-norm}) can be written as
\be
&{}&
\sum_\sigma
\sum_{k_1\dots k_m}
\overline{f_1(k_1)}\dots
\overline{f_m(k_m)}
g_{\sigma(1)}(k_1)
\dots
g_{\sigma(m)}(k_m)
\frac{1}{N^m}
\sum_{A\dots Z=1}^N
\underbrace{
\langle O|\dots\langle O|}_N
A_{k_1}\dots Z_{k_m}
\underbrace{
|O\rangle\dots |O\rangle}_N\label{az}
\ee
Since $m$ is fixed and we are interested in the limit $N\to\infty$ we
can assume that $N>m$. Each element of the sum over
$A_{k_1}\dots Z_{k_m}$ in (\ref{az}) can be associated with a unique
point $(A,\dots,Z)$ in an $m$-dimensional lattice embedded in a cube
with edges of length $N$.

Of particular interest are those points of the cube, the
coordinates of which are all different. Let us denote the subset of
such points by
$C_0$. For $(A,\dots,Z)\in C_0$
\be
\underbrace{
\langle O|\dots\langle O|}_N
A_{k_1}\dots Z_{k_m}
\underbrace{
|O\rangle\dots |O\rangle}_N
=
Z(\bbox k_1)\dots Z(\bbox k_m)\label{contr}
\ee
no matter what $N$ one considers and what are the numerical
components in $(A,\dots,Z)$. (This makes sense only for $N\geq m$; otherwise
$C_0$ would be empty). Therefore each element of $C_0$ produces an
identical contribution (\ref{contr}) to (\ref{az}). Let us denote the
number of points in $C_0$ by $N_0$.

The sum (\ref{az}) can be now written as
\be
{}&{}&
\sum_\sigma
\sum_{k_1\dots k_m}
\overline{f_1(k_1)}\dots
\overline{f_m(k_m)}
g_{\sigma(1)}(k_1)
\dots
g_{\sigma(m)}(k_m)
{\cal P}_0
Z(\bbox k_1)\dots Z(\bbox k_m)\nonumber\\
&{}&\pp{==}
+
\sum_\sigma
\sum_{k_1\dots k_m}
\overline{f_1(k_1)}\dots
\overline{f_m(k_m)}
g_{\sigma(1)}(k_1)
\dots
g_{\sigma(m)}(k_m)
\frac{1}{N^m}
\sum_{(A\dots Z)\notin C_0}
\underbrace{
\langle O|\dots\langle O|}_N
A_{k_1}\dots Z_{k_m}
\underbrace{
|O\rangle\dots |O\rangle}_N.\label{az'}
\ee
The coefficient ${\cal P}_0=\frac{N_0}{N^m}$ represents a probability
of $C_0$ in the cube.
The elements of the remaining sum over $(A\dots Z)\notin C_0$ can be also
grouped into classes according to the values of
$
\langle O|\dots\langle O|
A_{k_1}\dots Z_{k_m}
|O\rangle\dots |O\rangle$. There are $m-1$ such different classes,
each class has its associated probability ${\cal P}_j$, $0<j\leq
m-1$, which will appear  in the sum in an analogous role as
${\cal P}_0$.

The proof is completed by the observation that
\be
\lim_{N\to\infty}{\cal P}_0 &=& 1,\\
\lim_{N\to\infty}{\cal P}_j &=& 0,\quad 0<j.
\ee
Indeed, the probabilities are unchanged if one rescales the cube to
$[0,1]^m$. The probabilities are computed by means of an
$m$-dimensional uniformly distributed measure.
$N\to\infty$ corresponds to the continuum
limit, and in this limit the sets of points of which at least two
coordinates are equal are of $m$-dimensional measure zero.
\rule{5pt}{5pt}
\medskip

\noindent
{\it Comments:\/} (a) The thermodynamic limit is naturally equipped
with the scalar product yielding orthogonality relation of the form
(\ref{delta}). However, for small $N$ there will be differences if $m$
is large. On the other hand if $N$ is sufficiently large then the values
of $m$ for which the corrections are non-negligible must be also
large. But then a classical limit will be justified and the use of
non-canonical coherent states should again give the correct
description. (b) Concrete values of ${\cal P}_j$ for some small $m$
were given in \cite{I}.
For $m=2$: ${\cal P}_0=1-1/N$,  ${\cal P}_1=1/N$;
for $m=3$: ${\cal P}_0=1-3/N+2/N^2$,  ${\cal P}_1=3/N-3/N^2$,
${\cal P}_2=1/N^2$. In general we do not have to assume that $p_N=1$.
If $p_N$ are general probabilities then the coefficients involve
averages. For $m=2$: ${\cal P}_0=1-\langle 1/N\rangle$, ${\cal
P}_1=\langle 1/N\rangle$;
for $m=3$: ${\cal P}_0=1-\langle 3/N\rangle+\langle 2/N^2\rangle$,
${\cal P}_1=\langle 3/N\rangle-\langle 3/N^2\rangle$,
${\cal P}_2=\langle 1/N^2\rangle$, where $\langle 1/N\rangle=\sum_N p_N/N$
etc. The normalization in terms of $\langle\cdot|\cdot\rangle_Z$ is
then obtained under the assumption that all those averages vanish,
which can hold only approximately, meaning that the probability $p_N$
is peaked in a region of large $N$s.

\subsection{States generated by field operators and a first hint
indicating absence of ultraviolet divergences}

Consider the single-oscillator vector potential operator $A_a(x)$
which is related to (\ref{A_a}) by $\uu
A_a(x)=\oplus_{\frac{1}{\sqrt{n}}} A_a(x)$ and acting with $A_a(x)$ on a
single-oscillator vacuum $|O\rangle$ define
the vector
\be
|{A}_a(x)\rangle
&=&
{A}_a(x)|O\rangle\\
&=&
-i\int d\Gamma(\bbox k)e^{ik\cdot x}\Big(
m_{a}(\bbox k)
a(\bbox k,-)^{\dag}
+
\bar m_{a}(\bbox k)
a(\bbox k,+)^{\dag}\Big)
|O\rangle\\
&=&
-i\int d\Gamma(\bbox k)e^{ik\cdot x}O_M(\bbox k)|\bbox k\rangle\Big(
m_{a}(\bbox k)
a_-^{\dag}|0,0\rangle
+
\bar m_{a}(\bbox k)
a_+^{\dag}|0,0\rangle\Big)
\ee
Its multi-particle analogue is
\be
|{\uu A}_a(x)\rangle
&=&
{\uu A}_a(x)|\uu O\rangle\\
&=&
\oplus_{N=1}^\infty
\sqrt{\frac{p_N}{N}}
\Big(
|{A}_a(x)\rangle
\underbrace{
|O\rangle\dots|O\rangle}_{N-1}
+\dots+
\underbrace{
|O\rangle\dots|O\rangle}_{N-1}
|{A}_a(x)\rangle
\Big).
\ee
The positive definite scalar product
\be
\langle{\uu A}_a(y)|
(-g^{ab})
|{\uu A}_b(x)\rangle
&=&
\langle{A}_a(y)|
(-g^{ab})
|{A}_b(x)\rangle
=
2 \int d \Gamma(\bbox k) e^{ik\cdot (x-y)}Z(\bbox k) \label{UV}
\ee
shows that there is no ultraviolet divergence at $x=y$ since
$\int d \Gamma(\bbox k)Z(\bbox k)=1$.

It is easy to understand that the same property will hold also for general
states. To see this let us write the single-oscillator field
operator as a function of the operator
$\hat k_a=\int d\Gamma(\bbox k)k_a|\bbox k\rangle\langle\bbox k|$,
i.e.
\be
{A}_a(x)
&=&
im_{a}(\hat{\bbox k})
\big(e^{-i\hat k\cdot x}\otimes a_+
-e^{i\hat k\cdot x}\otimes a_-^{\dag}\big)
+
i\bar m_{a}(\hat{\bbox k})
\big(
e^{-i\hat k\cdot x}\otimes a_-
-e^{i\hat k\cdot x}\otimes a_+^{\dag}\big).
\ee
The operators $m_{a}(\hat{\bbox k})$ and $\bar m_{a}(\hat{\bbox k})$
are functions of the operator $\hat {\bbox k}$ and are defined in the
standard way via the spectral theorem. Moreover, they are complex
combinations of two {\it bounded\/} operators representing directions
of transverse polarizations. The remaining operators ($a_j$,
$a_j^{\dag}$, and $e^{\pm i\hat k\cdot x}$) are also well
behaved. Particularly striking is the fact that the {\it distribution\/}
$\int d\Gamma(\bbox k)e^{ik\cdot x}$ is replaced by the {\it unitary\/}
operator
\be
e^{i\hat k\cdot x}
=
\int d\Gamma(\bbox k)e^{ik\cdot x}|\bbox k\rangle\langle\bbox k|.
\ee
The latter property is at the very heart of the regularities encountered
in the non-canonical formalism.

Now, it is widely
known that configuration-space renormalization of ultraviolet
divergences can be reduced
to an appropriate treatment of products of field operators on the
diagonals $x=y$ \cite{Scharf,Connes}.
The formula (\ref{UV}) is a strong indication that
such divergences may be absent in the non-canonical framework.
Actually, the analysis of perturbation theory in nonrelativistic
quantum optics given in \cite{I} and further elaborated in Sec. X of
the present paper, shows that $Z(\bbox k)$
occurs in exactly those places where ultraviolet form-factors are
expected to appear.
The same property will
hold for non-canonically quantized fermionic fields.

\section{Statistics of excitations}

It is an experimental fact that laser beams produce Poissonian
statistics of photocounts. At the theoretical level of canonical
quantum optics the Poisson
distribution follows trivially from the form of canonical coherent
states. In the non-canonical case the exact Poisson statistics is
characteristic of the single-oscillator ($N=1$) sector. For $1<N<\infty$ the
statistics of excitations is non-Poissonian. At the other extreme
is the thermodynamic limit for multi-oscillator states.
In what follows we will show that in
the limit $N\to\infty$ one recovers {\it the same\/} Poisson distribution as
for $N=1$. This, at a first glance unexpected, result justifying our
definitions in terms of displacement operators is a consequence of
certain classical universality properties of the Poisson distribution.

We will also
return to the question of thermal states and the Planck formula. In
\cite{I} it was argued that non-CCR quantization implies deviations
from the black-body law. However, a consistent interepretation of the
field in terms of the thermodynamic limit shows that no deviations
should be expected.

\subsection{Multi-oscillator coherent states}

To study the thermodynamic limit of multi-oscillator coherent states
we simplify the discussion by taking an exactly $N$-oscillator
coherent state (\ref{mocs}) ($N\gg 1$ is fixed and $p_N=1$), i.e.
\be
|\uu O_\alpha\rangle
&=&
\underbrace{|O_{\alpha_N}\rangle\otimes\dots
\otimes |O_{\alpha_N}\rangle}_N
\ee
where
\be
|O_{\alpha_N}\rangle
&=&
\int d\Gamma(\bbox k)O(\bbox {k})
|\bbox {k},s\rangle|
\alpha (\bbox {k},+)/\sqrt{N}\rangle|\alpha (\bbox {k},-)/\sqrt{N}\rangle.
\ee
The average number of excitations in this state is
\be
\langle n\rangle
&=&
\sum_{s}\int d\Gamma(\bbox k)Z(\bbox {k})
|\alpha (\bbox {k},s)|^2.
\ee
The simplest case is the one where $\alpha (\bbox {k},s)=\alpha={\rm
const}$. Then $\langle n\rangle=|\alpha|^2$ and the statistics of
excitations of single-oscillator coherent states $|O_{\alpha_N}\rangle$
is Poissonian with the distribution
$p_n=e^{-|\alpha_N|^2}|\alpha_N|^{2n}/n!$

$m$ excitations distributed in the ensemble of $N$ oscillators can be
represented by the ordered $m$-tuple $(j_1,\dots,j_m)$,
$1\le j_1\le\dots\le j_m \le N$.  For
example, for $m=10$, $N=12$, the point $(2,2,2,5,5,7,7,7,11,11)$
represents 10 excitations distributed in the
ensemble of 12 oscillators as follows: 3 excitations in 2nd
oscillator, 2 in the 5th one, 3 in the 7th, and 2 in the 11th. Such
points form a subset of the cube $[0,N]^m$, the interior of the set
corresponding to points whose all the indices are different. The latter
means that the interior represents situations where there are $m$
oscillators excited, and each of them is in the first excited state.
The boundary of this set consists of points representing at least one
oscillator in a higher excited state.
Probabilities of events represented by points with
the same numbers of repeated indices must be identical due to
symmetries.
Intuitively, the Poissonian
statistics of the thermodynamic limit follows
from the fact that the probability of finding a point belonging to
the boundary tends to zero as $N$ increases. The statistics is
dominated by Bernoulli-type processes with probabilities
related to the two lowest energy levels of a single oscillator in a
coherent state.

To make the argument more formal we introduce the following notation:
\be
X_m^{(N)}
&=&
\{x\in {\bbox N}^m; m\geq 1,x=(j_1,\dots,j_m), 1\le j_1\le\dots\le j_m \le
N\}
\nonumber\\
X_{n_1\dots n_k}^{(N)}
&=&
\{x\in X_m^{(N)};
x=(\underbrace{i_1,\dots,i_1}_{n_1},
\dots
\underbrace{i_k,\dots,i_k}_{n_k}),
i_1<\dots <i_k
\}
\nonumber\\
Y_m^{(N)}
&=&
\bigcup_{(n_1\dots n_k)\neq (1\dots 1)}X_{n_1\dots n_k}^{(N)}
\ee
If we add a single-element set $X_0^{(N)}$ containing the event
representing $N$ oscillators in their ground states we can represent
the set of all the events by the disjoint sum
\be
X^{(N)}
&=&
\bigcup_{m=0}^\infty X_m^{(N)}
\ee
The probability of finding the partition $m=n_1+\dots +n_k$ is
\be
P(X_{n_1\dots n_k}^{(N)})&=&
N_{n_1\dots n_k}p_{n_1}\dots p_{n_k}p_0^{N-k}
\nonumber\\
&=&
N_{n_1\dots n_k}\frac{e^{-N|\alpha_N|^2}|\alpha_N|^{2m}}{n_1!\dots n_k!}
\ee
where $N_{n_1\dots n_k}$ is the number of elements of $X_{n_1\dots
n_k}^{(N)}\subset X_{m}^{(N)}$.  The probability
that $m$ excitations are found is
\be
P(X_{m}^{(N)})&=&\sum_{n_1\dots n_k} P(X_{n_1\dots n_k}^{(N)}),
\ee
the sum being over all the partitions of $m$. Denote by
$
P(Y_{m}^{(N)}|X_{m}^{(N)})
$
the conditional probability of finding at least one oscillator in the
2nd or higher excited state under the condition that the sum of excitations
is $m>1$. We first prove the following

\medskip
\noindent
{\bf Lemma 1.}
\be
\lim_{N\to\infty} P(Y_{m}^{(N)}|X_{m}^{(N)})=0.
\ee
{\it Proof\/}: Since $Y_{m}^{(N)}\cap X_{m}^{(N)}=Y_{m}^{(N)}$ one
finds
\be
P(Y_{m}^{(N)}|X_{m}^{(N)})
&=&
\frac{\sum_{(n_1\dots n_k)\neq (1\dots 1)} P(X_{n_1\dots n_k}^{(N)})}
{\sum_{n_1\dots n_k} P(X_{n_1\dots n_k}^{(N)})}
\nonumber\\
&=&
\Big[1+N_{1\dots 1}\Big(
\sum_{(n_1\dots n_k)\neq (1\dots 1)} \frac{N_{n_1\dots n_k}}
{n_1!\dots n_k!}\Big)^{-1}\Big]^{-1}
\nonumber\\
&<&
\Big[1+N_{1\dots 1}\Big(
\sum_{(n_1\dots n_k)\neq (1\dots 1)} N_{n_1\dots n_k}
\Big)^{-1}\Big]^{-1}
\ee
However,
\be
\lim_{N\to\infty}
\frac{\sum_{(n_1\dots n_k)\neq (1\dots 1)} N_{n_1\dots n_k}}
{N_{1\dots 1}}=0
\ee
on the basis of the geometric argument we gave in the previous
section. This ends the proof. \rule{5pt}{5pt}

The main result of this section is the following version of the well
known Poisson theorem:

\medskip
\noindent
{\bf Theorem 2.}
Assume that $\alpha(\bbox k,s)=\alpha={\rm const}$. Then
\be
\lim_{N\to\infty} P(X_{m}^{(N)})
=
\frac{e^{-|\alpha|^2}|\alpha|^{2m}}{m!}.
\ee
{\it Proof\/}:
As an immediate consequence of the lemma we find
\be
\lim_{N\to\infty} P(X_{m}^{(N)})&=&\lim_{N\to\infty}
P(X_{m}^{(N)}-Y_{m}^{(N)})
\ee
which means that in the thermodynamic limit we can treat excitations of the
oscillators to 2nd and higher excited levels as events whose
probability is zero. The probabilities of ground and first excited states
follow from the single-oscillator Poisson distributions but
conditioned by the fact that only the lowest two levels are taken
into account.

We thus arrive at the standard Poisson process with
\be
P_N &=& \frac{p_1}{p_0+p_1}=\frac{|\alpha_N|^2}{1+|\alpha_N|^2}
=
\frac{|\alpha|^2/N}{1+|\alpha|^2/N}
\ee
and $\lim_{N\to\infty}NP_N=|\alpha|^2$. \rule{5pt}{5pt}

\subsection{Thermal states}

A single-oscillator free-field Hamiltonian $H$ has the usual eigenvalues
\be
E(\omega,n)=\hbar\omega \Big(n+\frac{1}{2}\Big).
\ee
The eigenvalues of the free-field Hamiltonian $\uu H$ at the
multi-oscillator level are sums of the
single-oscillator ones. In \cite{I} it was assumed that the
Boltzmann-Gibbs distribution of thermal radiation should be
constructed in terms of $\uu H$. Let us note, however, that such a
construction makes use of $\uu H$ as if it was a Hamiltonian of a
single element of a statistical ensemble. The discussion of the
thermodynamic limit we have given above, as well as the results of
\cite{II}, suggest that $\uu H$ is the Hamiltonian of the entire
ensemble of systems described by $H$, and it is $H$ and not $\uu H$
which should be used in the Boltzmann-Gibbs distribution. Then, of
course, the result will be the standard one and no deviations from
the Planck formula will occur.

\section{Commutators of 4-potentials and locality}

A straightforward calculation shows that the multi-oscillator vector
potential operator satisfies, for any space-time points $x$, $y$,
$z$, the commutators
\be
[{\uu A}_a(x),{\uu A}_b(y)]
&=&
\int d\Gamma(\bbox k)\uu I_{k}
\Big(
m_{a}(\bbox k)\bar m_b(\bbox k)e^{ik\cdot (y-x)}
-
\bar m_a(\bbox k)m_b(\bbox k)e^{ik\cdot (x-y)}
\Big)\nonumber
\\
&\pp =&+
\int d\Gamma(\bbox k)\uu I_{k}
\Big(
\bar m_a(\bbox k)m_b(\bbox k)e^{ik\cdot (y-x)}
-
m_{a}(\bbox k)\bar m_b(\bbox k)e^{ik\cdot (x-y)}
\Big)\label{[A,A]}\\
\big[[{\uu A}_a(x),{\uu A}_b(y)],{\uu A}_c(z)\big]
&=&0.
\ee
To obtain more insight as to the meaning of the commutator
(\ref{[A,A]}) consider its coherent-state average evaluated in a state
of the form (\ref{mocs}):
\be
\langle \uu O_\alpha|
[{\uu A}_a(x),{\uu A}_b(y)]
|\uu O_\alpha\rangle
&=&
\int d\Gamma(\bbox k)Z(\bbox k)
\Big(
m_{a}(\bbox k)\bar m_b(\bbox k)e^{ik\cdot (y-x)}
-
\bar m_a(\bbox k)m_b(\bbox k)e^{ik\cdot (x-y)}
\Big)
\\
&\pp =&+
\int d\Gamma(\bbox k)Z(\bbox k)
\Big(
\bar m_a(\bbox k)m_b(\bbox k)e^{ik\cdot (y-x)}
-
m_{a}(\bbox k)\bar m_b(\bbox k)e^{ik\cdot (x-y)}
\Big).
\ee
The Minkowski-space metric tensor can be decomposed
\cite{PR} in terms of null tetrads as follows
\be
g_{ab}
&=&
k_a\omega_b+\omega_a k_b-m_a\bar m_b-m_b\bar m_a.\label{g-id}
\ee
With the help of this identity we can write
\be
\langle \uu O_\alpha|
[{\uu A}_a(x),{\uu A}_b(y)]
|\uu O_\alpha\rangle
&=&
\int d\Gamma(\bbox k)Z(\bbox k)
\big(k_{a}\omega_b(\bbox k)
+
k_{b}\omega_a(\bbox k)
\big)
\big(
e^{ik\cdot (y-x)}
-
e^{ik\cdot (x-y)}
\big)\nonumber\\
&\pp =&+g_{ab}
\int d\Gamma(\bbox k)Z(\bbox k)
\big(
e^{ik\cdot (x-y)}
-
e^{ik\cdot (y-x)}
\big)\label{<A>}
\ee
It is known that terms such as the first integral vanish if the
potential couples to a conserved current. The same property
guarantees gauge independence of the formalism. Therefore we can
concentrate only on the explicitly gauge independent term
proportional to $g_{ab}$.
Denote
\be
D_Z(x)
&=&
i\int d\Gamma(\bbox k)Z(\bbox k)
\big(
e^{-ik\cdot x}
-
e^{ik\cdot x}
\big)
\ee
For $Z(\bbox k)={\rm const}=Z$ we get $D_Z(x)$ proportional to the
Jordan-Pauli function,
\be
D_{Z}(x)=Z\, D(x)
\ee
which vanishes for spacelike $x$.
However, the choice of
constant $Z(\bbox k)$ is excluded by the requirement of
square-integrability of $O$. Therefore the requirement that the
vacuum state be square-integrable seems to introduce some kind of
nonlocality into the formalism.

There are two possibilities one can contemplate. First of all, one
can perform the calculations with arbitrary $O$ and then perform a
renormalization (we have seen that such a step is necessary even in
the nonrelativistic case). After the renormalization we can go to the
``flat" pointwise limit $Z(\bbox k)\to 0$, $\parallel O\parallel=1$,
corresponding to the uniform
distribution of all the frequencies. Second, performing the
calculations in a preferred frame we can consider equal-time
commutation relations
\be
\langle \uu O_\alpha|
[{\uu A}_a(t,\bbox x),{\uu A}_b(t,\bbox y)]
|\uu O_\alpha\rangle
&=&
\int d\Gamma(\bbox k)Z(\bbox k)
\big(k_{a}\omega_b(\bbox k)
+
k_{b}\omega_a(\bbox k)
\big)
\big(
e^{i\vec k\cdot (\vec y-\vec x)}
-
e^{i\vec k\cdot (\vec x-\vec y)}
\big)\nonumber\\
&\pp =&+g_{ab}
\int d\Gamma(\bbox k)Z(\bbox k)
\big(
e^{i\vec k\cdot (\vec x-\vec y)}
-
e^{i\vec k\cdot (\vec y-\vec x)}
\big)
\ee
The last term will vanish if
\be
Z(\bbox k)=Z(-\bbox k)
\ee
i.e. if the vacuum is 3-inversion invariant.
This can hold, however only in one reference frame unless $O$ is
constant, which we exclude.

One can conclude that non-canonically quantized electrodynamics
is not a local quantum field theory, at least in the strict standard
sense. This is not very surprizing since
the presence of $Z(\bbox k)$ in the integrals introduces some kinds of
effective extended structures, a consequence of nontrivial structures
of non-canonical vacua.
The issue requires further studies. In particular, it is
important to understand an influence of the thermodynamic limit
$N\to\infty$ on locality problems in the context of relativistic
perturbation theory.

There is an intriguing analogy between the kind of non-locality we
have obtained and the one encountered in field theory in
non-commutative space-time \cite{Jackiw,Cai,geometry}.

\section{Radiation fields associated with classical currents}

The problem of radiation fields is interesting for several reasons.
First of all, the radiation fields satisfy homogeneous Maxwell
equations so that the theory we have developed
can be directly applied. Second, this is one of the simplest ways of
addressing the question of infrared divergences within the
non-canonical framework.

It is widely known \cite{IZBB,BD} that  in the
canonical theory the scattering matrix corresponding
to radiation fields produced by a classical transversal
current is given, up to a phase, by a
coherent-state displacement operator
$e^{-i\int d^4y J(y)\cdot A_{\rm in}(y)}$.
One of the consequences of such
an approach is the Poissonian statistics of photons emitted by
classical currents. An unwanted by-product of the construction is the infrared
catastrophe.

Starting with the same $S$ matrix but expressed in terms of
non-canonical ``in" fields we obtain the non-canonical displacement
operator.
Photon statistics is Poissonian in the thermodynamic limit but the infrared
catastrophe is automatically eliminated.

\subsection{Classical radiation field}

Let us assume that we deal with a classical transversal current $J_a(x)$
whose Fourier transform is $\tilde J_a(k)=\int d^4x e^{ik\cdot x}J_a(x)$.
Transversality means here that
\be
\tilde J_a(|\bbox k|,\bbox k)=
\bar m_a(\bbox k)\tilde J_{10'}(|\bbox k|,\bbox k)
+
m_a(\bbox k)\tilde J_{01'}(|\bbox k|,\bbox k).
\ee
Formally, a solution of Maxwell equations
\be
\Box{\uu A}_a(x)=J_a(x)\uu I
\ee
can be written as
\be
{\uu A}_a(x)
&=&
{\uu A}_{a\rm in}(x)+\int d^4y D_{\rm ret}(x-y)J_a(y)\uu I\\
&=&
{\uu A}_{a\rm out}(x)+\int d^4y D_{\rm adv}(x-y)J_a(y)\uu I.
\ee
Here ${\uu A}_{a\rm in}$ and ${\uu A}_{a\rm out}$ are solutions of
homogeneous equations. $D_{\rm ret}$ and $D_{\rm adv}$ are the
retarded and advanced
Green functions whose difference is the Jordan-Pauli function
\be
D(x)=i\int d\Gamma(\bbox k)\big(e^{-ik\cdot x}-e^{ik\cdot x}\big).
\ee
The 4-potential of the radiation field is
\be
{\uu A}_{a\rm rad}(x)
&=&
{\uu A}_{a\rm out}(x)
-
{\uu A}_{a\rm in}(x)=\int d^4y D(x-y)J_a(y)\uu I
\ee
and leads to the field spinors
\be
\uu\varphi_{XY\rm rad}(x)
&=&
\int d\Gamma(\bbox k)
\Big(\pi{_{(X}}\bar\pi{^{Y'}} \tilde J_{Y)Y'}(k)e^{-ik\cdot x}
+\pi{_{(X}}\bar\pi{^{Y'}}\overline{\tilde J}_{Y)Y'}(k)e^{ik\cdot x} \Big)
\uu I\\
{\uu{\bar\varphi}}_{X'Y'\rm rad}(x)
&=&
\int d\Gamma(\bbox k)
\Big(\bar\pi{_{(X'|}}\pi{^{Y}} \overline{\tilde J}_{Y|Y')}(k)e^{ik\cdot x}
+\bar\pi{_{(X'|}}\pi{^{Y}}\tilde J_{Y|Y')}(k)e^{-ik\cdot x} \Big)
\uu I.
\ee
Comparing these formulas with expressions (\ref{varphi}) and
(\ref{varphi'}) valid for all solutions of free Maxwell equations one
finds
\be
f(\bbox k,+)
&=&
-\bar m^a(\bbox k)\tilde J_{a}(|\bbox k|,\bbox k)
=\tilde J_{01'}(|\bbox k|,\bbox k)=j(\bbox k,+)\label{J01}\\
f(\bbox k,-)
&=&
-m^a(\bbox k)\tilde J_{a}(|\bbox k|,\bbox k)
=\tilde J_{10'}(|\bbox k|,\bbox k)=j(\bbox k,-)\label{J10}\\
\uu a(\bbox k,s)_{\rm out}
&=&
\uu a(\bbox k,s)_{\rm in}+
j(\bbox k,s)\uu I.\label{in-out}
\ee
\subsection{Non-canonical radiation field}

Formula (\ref{in-out}) is analogous to the one from the canonical
theory. It is clear that although $\uu a(\bbox k,s)_{\rm in}$ and
$\uu a(\bbox k,s)_{\rm out}$ cannot be simultaneously of the form given by
(\ref{53}) and (\ref{58}), they do satisfy the non-CCR algebra
(\ref{65}). In spite of this the result (\ref{in-out}) is
not very satisfactory. Indeed, one expects that the scattering matrix
describing a non-canonical quantum field interacting with a classical
current is
\be
S=e^{i\phi}e^{-i\int d^4y J(y)\cdot\uu A_{\rm in}(y)}\label{S-m}
\ee
with some phase $\phi$. Then
\be
\uu A_{a\rm out}(x)
&=&
S^{\dag}\uu A_{a\rm in}(x)S\label{S-matrix}\\
&=&
\uu A_{a\rm in}(x)
-i\int d^4y J^{b}(y)
[\uu A_{a\rm in}(x),\uu A_{b\rm in}(y)]\label{Arad}.
\ee
Employing (\ref{[A,A]}), (\ref{J01}), (\ref{J10}) one can write
\be
\uu A_{a\rm rad}(x)
&=&
i\int d^4y J^{b}(y) \int d\Gamma(\bbox k)
\uu I_{\bbox k}
\Big(
\big(
e^{ik\cdot (x-y)}\bar m_am_b
-e^{-ik\cdot (x-y)}m_a\bar m_b
\big)
+
\big(
e^{ik\cdot (x-y)}m_a\bar m_b
-e^{-ik\cdot (x-y)}\bar m_am_b
\big)
\Big)\\
&=&
i\int d\Gamma(\bbox k)
\uu I_{\bbox k}\Big(m_a\big(e^{-ik\cdot x}j(\bbox k,+)
-e^{ik\cdot x}\overline{j(\bbox k,-)}\big)
+
\bar m_a\big(e^{-ik\cdot x}j(\bbox k,-)
-e^{ik\cdot x}
\overline{j(\bbox k,+)}\big)
\Big),
\ee
where $m_a=m_a(\bbox k)$, and
\be
\uu a(\bbox k,s)_{\rm out}
&=&
\uu a(\bbox k,s)_{\rm in}+
j(\bbox k,s)\uu I_{\bbox k}\label{in-out'}\\
&=&
\uu {\cal D}(j)^{\dag}
\uu a(\bbox k,s)_{\rm in}
\uu {\cal D}(j).
\ee
Consequently, the $S$ matrix is in the non-canonical theory
proportional to the {\it non-canonical\/} displacement operator
\be
S=e^{i\phi}\uu {\cal D}(j).
\ee
This fact will be shown to eliminate the infrared catastrophe.

\subsection{Propagators}

Evaluating the average of (\ref{Arad}) in a coherent state
$|\uu O_\alpha\rangle$ one finds
\be
\langle\uu O_\alpha|
\uu A_{a\rm rad}(x)
|\uu O_\alpha\rangle
&=&
\int d^4y D_Z(x-y) J_a(y)+{\rm gauge \,term}.\label{+gauge}
\ee
The irrelevant gauge term is a remainder of the first part of
(\ref{<A>}).  As expected the radiation field does not depend on what
$\alpha$ one takes in the coherent state, but does depend on the
vacuum structure. The presence of the regularized function
$D_Z(x-y)$ instead of $D(x-y)$
implies that the radiation signal propagates in a neighborhood of the
light cone. Any deviation from $c$ in velocity of signal propagation
can be regarded as an indication of a non-constant vacuum wave function
$O(\bbox k)$. To have a feel of the scale of the nonlocality assume
that $O(\bbox k)$ is constant up to, roughly, the Planck scale.
Corrections of the order of the classical electron radius can be seen
in the distance travelled by light if the photon travels
100 light years. Moreover, even in the orthodox canonical quantum
electrodynamics a detailed analysis of signal propagation leads to
small deviations from velocity of light, especially at small
distances \cite{Man}. It may be difficult to experimentally
distinguish between the two effects. A similar effect was predicted
for Maxwell fields in non-commutative space-time \cite{Jackiw,Cai}.

Using (\ref{g-id}) one can rewrite (\ref{[A,A]}) as
\be
[{\uu A}_a(x),{\uu A}_b(y)]
&=&
g_{ab}
\int d\Gamma(\bbox k)\uu I_{\bbox k}
\Big(e^{ik\cdot (x-y)}-e^{ik\cdot (y-x)}\Big)
+\dots\label{hat D}
\ee
where the dots stand for all the terms which are gauge dependent and
do not contribute to physically meaningful quantities.
We can therefore identify
\be
\uu D(x)=i\int d\Gamma(\bbox k)\uu I_{\bbox k}
\Big(e^{-ik\cdot x}-e^{ik\cdot x}\Big)
\ee
as the operator responsible for the appearence of the smeared out
Jordan-Pauli function $D_Z$ in the
coherent-state average (\ref{+gauge}). $\uu D(x)$ is a
translation-invariant scalar-field operator solution of the
d'Alembert equation, i.e.
\be
\Box \uu D(x)&=&0,\label{d'A}\\
\uu V_{\Lambda,y}^{\dag}\uu D(x) \uu V_{\Lambda,y}&=&\uu
D(\Lambda^{-1}x).
\ee
The operator analogues
of retarded and advanced Green functions are
\be
\uu D_{\rm ret}(x)
&=&
\Theta(x_0)\uu D(x),\\
\uu D_{\rm adv}(x)
&=&
-\Theta(-x_0)\uu D(x),\\
\uu D(x)&=&\uu D_{\rm ret}(x)-\uu D_{\rm adv}(x).
\ee
Eq. (\ref{d'A}) implies that the operators
\be
\int d^4y \uu D_{\rm ret}(x-y)J_a(y)
\ee
and
\be
\int d^4y \uu D_{\rm adv}(x-y)J_a(y)
\ee
differ by at most a solution of the homogeneous equation
\be
\Box \uu A_a(x)=0. \label{hom}
\ee
(\ref{d'A}) implies also that one can {\it define\/}
\be
\uu \delta(x):=\Box \uu D_{\rm adv}(x)=\Box \uu D_{\rm ret}(x).
\ee
It follows that having a solution ${\uu A}_{a\rm in}(x)$ of
(\ref{hom}) one can define another solution ${\uu A}_{a\rm out}(x)$ of
(\ref{hom}) by means of
\be
{\uu A}_a(x)
&=&
{\uu A}_{a\rm in}(x)+\int d^4y \uu D_{\rm ret}(x-y)J_a(y)\\
&=&
{\uu A}_{a\rm out}(x)+\int d^4y \uu D_{\rm adv}(x-y)J_a(y),
\ee
simulatneously guaranteeing that the correct $S$-matrix conditions
(\ref{S-m}), (\ref{S-matrix}) are fulfilled up to, perhaps, a gauge
transformation. ${\uu A}_a(x)$ is a solution of
\be
\Box \uu A_a(x)=\uu J_a(x) \label{inhom}
\ee
where
\be
\uu J_a(x)&=&\int d^4y \uu \delta(x-y)J_a(y).
\ee
\subsection{The problem of infrared catastrophe}

To close this part of the discussion let us consider the issue of infrared
catastrophe. We have to compute an avarage number of photons in the
state $\uu D(j)|\uu O\rangle$. The number-of-photons operator is
\be
\uu n=\oplus_1 (1\otimes \sum_sa^{\dag}_sa_s).
\ee
The $1$ in the above formula is the identity in the $\bbox k$ space
and $a_s$ satisfy CCR.
The average reads
\be
\langle n\rangle
&=&
\langle \uu O_j|\uu n|\uu O_j\rangle
=
\sum_s\int d\Gamma(\bbox k)Z(\bbox k)|j(\bbox k,s)|^2.
\ee
The four-momentum of the radiation field is
\be
\langle P_a\rangle
&=&
\langle \uu O_j|\uu P_a|\uu O_j\rangle
=
\sum_s\int d\Gamma(\bbox k)k_a Z(\bbox k)|j(\bbox k,s)|^2.
\ee
$O(\bbox k)$ belongs to a carrier space of an
appropriate unitary representation of the Poincar\'e group. As such this is a
differentiable function vanishing
at the origin $k=0$ of the light cone. This is a consequence of the
fact that the cases $k=0$ and $k\neq 0$, $k^2=0$, correspond to
representations of the Poincar\'e group induced from $SL(2,C)$ and $E(2)$,
respectively (for another justification see \cite{Woodhouse}).

Hence, the regularization of the infrared divergence is implied by
relativistic properties of the field. It is quite remarkable that all
the divergences are regularized automatically by the
same property of the formalism: The nontrivial structure of the
vacuum state. In the case of ultraviolet and vacuum divergences the
regularization is a consequence of square integrability of $O$.

\section{Selected questions of physical interpretation}

In this section we want to address three problems which allow to
better understand on physical grounds the formal structures
characteristic of our choice of field operators. First we will
discuss in a simplified model (scalar field, discrete spectrum of
frequencies) the relationship between the non-CCR approach and the
canonical one. We will perform this in two versions: One using an
exact result, and the second one based on perturbation theory.
Making use of the same model we will interpret the factor
$1/\sqrt{N}$ occuring in the definition of multi-oscillator field
operators, and finally we will look more closely at the structure of
the state space.

\subsection{Canonical limit and the second hint indicating absence of
ultraviolet divergences: Exact example}
\label{example}

We now want to
make a link between the non-CCR theory with nonunique vacuum
and the one based on CCR and the unique vacuum. One may regard it as
a test of experimental consequences of the non-canonical approach.
The analysis of Poisson statistics of coherent states indicates that
the link may be provided by the thermodynamic limit $N\to\infty$. A
similar conclusion follows from the analysis of orthogonality
properties of multi-photon states.

In what follows we shall discuss this in more detail on a simplified
model of a non-CCR scalar field in dipole approximation
interacting with a two-level atom located at $\bbox x=0$.
It is known that without a regularization of the coupling parameters
by $g_{\bbox k}\to g_{\bbox k} Z_{\bbox k}^{\frac{1}{2}}$, with
$Z_{\bbox k}^{\frac{1}{2}}\to 0$ for
$|\bbox k|\to\infty$, the model is highly
ultraviolet divergent (even more than its relativistic counterpart).
On the other hand the vast majority of experimental tests of quantum
radiation fields is based on this type of calculation.

We will compare predictions of two models: The CCR one with
regularized coupling constants, and the non-CCR quantized field {\it
without\/} regularization of the Hamiltonian.

Consider the interaction-picture Hamiltonian
\be
H(t)&=&
i\frac{\hbar\alpha}{2}
\sum_{\bbox k}
\sqrt{\frac{\hbar}{2\omega_k}}\Big(
R_+a{_{\bbox k}}e^{-i\Delta_{\bbox k} t}
-
R_-a{_{\bbox k}}^{\dag}e^{i\Delta_{\bbox k} t}
\Big).
\ee
The coupling constant $\alpha$ depends linearly on the ratio
$e_0/\sqrt{V}$ (charge $e_0$ and quantization volume $V$).
We assume the CCR algebra
$[a_{\bbox k},a^{\dag}_{\bbox k'}]=\delta_{\bbox k\bbox k'}$ with the unique
vacuum $|0_{\rm ccr}\rangle$, and $R_\pm=|\pm\rangle\langle\mp|$.

Of particular interest is the vacuum-to-vacuum amplitude
\be
F(t)=\langle O|U(t)|O\rangle
\ee
which can be exactly computed in both models. In the CCR case,
taking $|O\rangle=|0_{\rm ccr}\rangle|+\rangle$, in order
to compute the amplitude the Hamiltonian
$H(t)$ has to be replaced by its regularized version
\be
H_{\rm reg}(t)&=&
i\frac{\hbar\alpha}{2}
\sum_{\bbox k} Z_{\bbox k}^{\frac{1}{2}}
\sqrt{\frac{\hbar}{2\omega_k}}\Big(
R_+a{_{\bbox k}}e^{-i\Delta_{\bbox k} t}
-
R_-a{_{\bbox k}}^{\dag}e^{i\Delta_{\bbox k} t}
\Big).
\ee
Evaluating the vacuum average of the Dyson expansion one finds the
Volterra-type integral equation
\be
F(t)&=&
1-C
\int_0^{t}dt_1\int_0^{t_1}dt_2
f_Z(t_1-t_2)F(t_2)\label{A2}
\ee
where $C=\alpha^{2}\hbar/8$ and
\be
f_Z(\tau)=\sum_{\bbox k} Z_{\bbox k}\frac{e^{-i\Delta_{\bbox
k}\tau}}{\omega_k}
\label{fOtau}.
\ee
The amplitude expressed in terms of the Laplace transform, with
integration along a contour located to the right of all the poles of
the integrand, is
\be
F(t)
&=&
\frac{1}{2\pi i}
\int_\Gamma
dz\frac{e^{zt}}{z+C
\sum_{\bbox k}Z_{\bbox k}\frac{1}{\omega_k}\frac{1}{i\Delta_{\omega_k}+z}}.
\label{A3}
\ee
Let us now take the interaction-picture Hamiltonian
\be
\uu H(t)&=&
i\frac{\hbar\alpha}{2}
\sum_{\bbox k}
\sqrt{\frac{\hbar}{2\omega_k}}\Big(
R_+\uu a{_{\bbox k}}e^{-i\Delta_{\bbox k} t}
-
R_-\uu a{_{\bbox k}}^{\dag}e^{i\Delta_{\bbox k} t}
\Big),\label{ipH}
\ee
with the non-CCR operators
$[\uu a_{\bbox k},\uu a^{\dag}_{\bbox k'}]=
\delta_{\bbox k\bbox k'}\uu I_{\bbox k}$, $\sum_{\bbox k} \uu
I_{\bbox k}=\uu I$,
and the non-canonical vacuum
\be
|\uu O\rangle
&=&
\bigoplus_{N=1}^\infty\sqrt{p_N}\sum_{k_1\dots k_N}
O_{k_1}\dots O_{k_N}
|\bbox k_1\dots \bbox k_N\rangle
\ee
which is of the form (\ref{uu O}).
Let us note that now we do not introduce any regularization of the
coupling parameters.

Setting $Z_k=|O_k|^2$ and denoting the non-canonical amplitude by
$F'$ one can write
\be
F'(t)
&=&
\sum_{N=1}^\infty p_N\sum_{k_1\dots k_N}
Z_{k_1}\dots Z_{k_N}
\langle \bbox k_1\dots\bbox k_N|U(t)
|\bbox k_1\dots\bbox k_N\rangle\\
&=&
\sum_{N=1}^\infty p_N\sum_{k_1\dots k_N}
Z_{k_1}\dots Z_{k_N}
F(t)_{k_1\dots k_N}.
\ee
The Dyson expansion
leads to a similar Volterra-type equation as before
\be
F(t)_{k_1\dots k_N}
&=&
1
-
\frac{C}{N}
\int_0^{t}dt_1\int_0^{t_1}dt_2
f(t_1-t_2)_{k_1\dots k_N}
F(t_2)_{k_1\dots k_N}
\label{int-eq}
\ee
with
\be
f(\tau)_{k_1\dots k_N}=\sum_{j=1}^N\frac{e^{-i\Delta_{\omega_{j}}\tau}
}{\omega_{j}},\label{ftau}
\ee
$\omega_j=\omega_{k_j}=|\bbox k_j|$,
and the same constant $C$. The solution is
\be
F(t)_{k_1\dots k_N}
=
\frac{1}{2\pi i}
\int_{\Gamma}
dz\frac{e^{zt}}
{z+\frac{C}{N} \sum_{j=1}^N\frac{1}{\omega_{j}}\frac{1}
{i\Delta_{\omega_{j}}+z}}
\label{F_{}}
\ee
where $\Gamma$ is any contour parallel to the imaginary axis and to
the right of all the poles of the integrand.

The solution (\ref{F_{}}) looks
very similar to (\ref{A3}) but there are also evident differences.
For example the sum in the denominator of the inegrand in
(\ref{F_{}}) is finite, involves no regularizations, and $C$ is
replaced by $C/N$.

To see the link between $F(t)$ and $F'(t)$ let us assume that there are
exactly $N$ oscillators (i.e. $p_N=1$ for some $N$ and zero
otherwise) and consider the thermodynamic limit $N\to \infty$. This
is the same type of reasoning we have employed in
discussion of the Poisson statistics of coherent states and
normalization of multi-photon states.

\medskip
\noindent
{\bf Theorem 3.}
Under the above assumptions
\be
\lim_{N\to\infty} F'(t)=F(t).
\ee
{\it Proof\/}:
The poles of the integrand in (\ref{F_{}}) are equal to the eigenvalues of
\be
\left(
\begin{array}{ccccc}
0 &
-\sqrt{\frac{C}{\omega_1 N}}
&
\dots
&

&
-\sqrt{\frac{C}{\omega_N N}}
\\
\sqrt{\frac{C}{\omega_1 N}}
&i\Delta_{\omega_1}&0&\dots&0\\
\vdots& & & & \\
\sqrt{\frac{C}{\omega_N N}}
&0& \dots &  &
i\Delta_{\omega_N}
\end{array}
\right)
\ee
and, hence, are purely imaginary. The parameters $C$, $N$ and
$\omega_1,\dots\omega_N$ in (\ref{F_{}}) are fixed, integration is
over any contour localized to the right of all the poles, and the poles are
imaginary. We assume that the spectrum of $\omega$s contains a
minimal $\omega>0$ (as in cavity). It follows that
the contour can be shifted sufficiently far to the right so that the
inequality
\be
\Big|\frac{C}{N}
\sum_{j=1}^N\frac{1}{\omega_j}\frac{1}{z}\frac{1}{i\Delta_{\omega_j}+z}\Big|<1
\ee
is satisfied for any choice of $\omega_1,\dots,\omega_N$, and
\be
\frac{1}{1+\frac{C}{N}
\sum_{j=1}^N\frac{1}{\omega_j}\frac{1}{z}\frac{1}{i\Delta_{\omega_j}+z}}
&=&
\sum_{n=0}^\infty \Big(
-\frac{C}{N}
\sum_{j=1}^N\frac{1}{\omega_j}\frac{1}{z}\frac{1}{i\Delta_{\omega_j}+z}\Big)^n
\ee
The amplitude of interest can be thus written as
\be
F'(t)
&=&
\frac{1}{2\pi i}
\sum_{k_1\dots k_N}
Z_{k_1}\dots Z_{k_N}
\sum_{n=0}^\infty (
-C)^n
\int_\Gamma
dz\frac{e^{zt}}{z^{n+1}}
\frac{1}{N^n}
\Big(\sum_{j=1}^N\frac{1}{\omega_j}\frac{1}{i\Delta_{\omega_j}+z}\Big)^n
\\
&=&
\frac{1}{2\pi i}
\sum_{k_1\dots k_N}
Z_{k_1}\dots Z_{k_N}
\sum_{n=0}^N (
-C)^n
\int_\Gamma
dz\frac{e^{zt}}{z^{n+1}}
\frac{1}{N^n}
\Big(\sum_{j=1}^N\frac{1}{\omega_j}\frac{1}{i\Delta_{\omega_j}+z}\Big)^n
\nonumber\\
&\pp =&+
\frac{1}{2\pi i}
\sum_{k_1\dots k_N}
Z_{k_1}\dots Z_{k_N}
F'_{n>N}(t)_{k_1\dots k_N}.
\ee
The convergence of the geometric series guarantees that for any
$\varepsilon$ there exists $N_\varepsilon$ such that
\be
|F'_{n>N_\varepsilon}(t)_{k_1\dots k_N}|<\varepsilon,
\quad
\Big|\sum_{k_1\dots k_{N_\varepsilon}}
Z_{k_1}\dots Z_{k_{N_\varepsilon}}
F'_{n>{N_\varepsilon}}(t)_{k_1\dots k_{N_\varepsilon}}\Big|
<\varepsilon.
\ee
Therefore
\be
\lim_{N\to\infty}F'(t)
&=&
\frac{1}{2\pi i}
\lim_{N\to\infty}
\sum_{n=0}^N (
-C)^n
\int_\Gamma
dz\frac{e^{zt}}{z^{n+1}}
\sum_{k_1\dots k_N}
Z_{ k_1}\dots Z_{ k_N}
\frac{1}{N^n}
\Big(\sum_{j=1}^N\frac{1}{ \omega_j}\frac{1}{i\Delta_{ \omega_j}+z}\Big)^n
\\
&=&
\frac{1}{2\pi i}
\lim_{N\to\infty}
\sum_{n=0}^N (
-C)^n
\int_\Gamma
dz\frac{e^{zt}}{z^{n+1}}
\Big[
\Big(\sum_{k}Z_{k}\frac{1}{\omega_k}
\frac{1}{i\Delta_{\omega_k}+z}\Big)^n
{\cal P}_0+\dots+
\sum_{k}Z_{k}\Big(\frac{1}{\omega_k}
\frac{1}{i\Delta_{\omega_k}+z}\Big)^n
{\cal P}_N\Big]\\
&=&
\frac{1}{2\pi i}
\lim_{N\to\infty}
\int_\Gamma
dz\frac{e^{zt}}{z}
\Big[
f_N^0
{\cal P}_0+\dots+
f_N^N
{\cal P}_N\Big],
\ee
where ${\cal P}_0\dots {\cal P}_N$ are the probabilities
employed in the proof of Theorem 1 and
$\lim_{N\to\infty}f_N^j<\infty$ for all $j$.
Using $\lim_{N\to\infty}{\cal P}_0=1$ and $\lim_{N\to\infty}{\cal
P}_j=0$ for $j>0$ we find
\be
\lim_{N\to\infty}F'(t)
&=&
\frac{1}{2\pi i}
\int_\Gamma
dz\frac{e^{zt}}{z+C
\sum_{\bbox k}Z_{\bbox k}\frac{1}{ \omega_k}\frac{1}{i\Delta_{ \omega_k}+z}}.
\ee
\rule{5pt}{5pt}
\medskip

It is important to note that $F(t)$ obtained as the limiting value of
$F'(t)$ contains $Z_k=|O_k|^2$ satisfying the
normalization condition $\sum_k Z_k=1$. Therefore although the
thermodynamic limit regularizes the {\it non-canonical\/}
amplitude as if it originated from the regularized {\it canonical\/}
Hamiltonian $H_{\rm reg}$, the regularization involves functions
$Z_k$ which are somewhat unusual from the viewpoint of
standard quantum optics.

Indeed, the typical choice one finds in the literature is of the form
\cite{RZ}
\be
Z_k(\Lambda)
&=&\frac{\Lambda}{\sqrt{\bbox k^2+\Lambda^2}}\approx
\left\{
\begin{array}{lll}
1 & {\rm for} & |\bbox k|\ll \Lambda\\
0 & {\rm for} & |\bbox k|\gg \Lambda
\end{array}
\right.
\ee
and it is essential for the standard interpretation that $Z_k\to 1$
with $\Lambda\to \infty$.
The regularization we arrive at must be square summable (or square
integrable) and normalized. So assume
\be
Z_k(\Lambda_1,\Lambda_2)=\left\{
\begin{array}{lll}
Z={\rm const} & {\rm for} &|\bbox k|\in [\Lambda_1, \Lambda_2]\\
{\rm tends\,to\,}0 & {\rm for} & |\bbox k|\notin [\Lambda_1, \Lambda_2]
\end{array}
\right.
\ee
But since $\sum_k Z_k=1$, the limit $\Lambda_2\to\infty$ is
accompanied by $Z\to 0$. An agreement with the standard formalism is
then obtained if one performs the latter limit under the constraint
$CZ=C_{\rm exp}$, i.e. renormalizes the coupling parameter $\alpha$
by $\alpha_{\rm exp}=\alpha Z^{\frac{1}{2}}$, which is essentially the
charge renormalization $e_{\rm exp}=e_0Z^{\frac{1}{2}}$.

All the examples studied so far suggest the following

\medskip
\noindent
{\bf Conjecture:} The thermodynamic limit $N\to\infty$
maps predictions of the non-canonical theory into those of the
canonical theory regularized by square-integrable formfactors.
Elimination of regularization must be
accompanied by charge renormalization.

\subsection{Canonical limit and the third hint suggesting absence of
ultraviolet divergences: Perturbation theory}
\label{example2}

Let us now discuss the thermodynamic limit of a perturbative
expansion to order $n$ of an arbitrary amplitude
\be
F_{fi}(t)
&=&
\langle
\uu \Psi_f|U(t)|\uu \Psi_i\rangle.
\ee
Examples of low order perturbation theory in concrete examples
were explicitly treated in \cite{I}. Here we want to discuss an
arbitrary order of a general amplitude corresponding to the
Hamiltonian (\ref{ipH}) which has led to the Conjecture from the
previous section.
It is sufficient to discuss the amplitude for the general multiphoton
states discussed in Theorem 1.

The interaction Hamiltonian (\ref{ipH}) is applicable to fields with discrete
frequency (i.e. cavity) spectrum. This is technically a simple case
since momentum eigenvectors are normalized in volume $V$
by the Kroenecker deltas.
Performing the expansion to order $n$ of the amplitude we arrive, for
each Feynman diagram in momentum space, at the term proportional to
\be
\langle \uu O|\uu a(f_1)\dots \uu a(f_m)
{\rm Mono}\big(\uu a(\bbox k_1),\dots,\uu a(\bbox k_n)\big)
\uu a(g_1)^{\dag}\dots\uu a(g_{m'})^{\dag}|\uu O\rangle\label{Mono}
\ee
where ${\rm Mono}(\dots)$ is a mononomial of order $n$ in creation
and annihilation operators. Due to the Kroenecker-delta normalization the term
(\ref{Mono}) vanishes {\it if and only if\/} an analogous term
does in the canonical theory (cf. the discusion in
\cite{I}). For this reason the non-CCR amplitude produces the
same Feynman diagrams, but the Feynman rules are modified by the
non-CCR algebra.

The latter modificatios are easy to identify. Simply, each time a
commutator of creation and annihilation operators produces a
delta in the standard theory, say $\delta_{kk'}$, here it is
replaced by $\delta_{kk'}\uu I_k$.
The Wick-theorem-type manipulations one has to perform under the vacuum
average are of the same type as those we have met in
the proof of Theorem 1 while normalizing multi-photon states.

The essential and final step follows from the main property of the
thermodynamic limit:
$\lim_{N\to\infty}{\cal P}_0=1$ and $\lim_{N\to\infty}{\cal P}_j=0$
for $j>0$. Therefore only those terms count in the limit which
have each vacuum average of the product $\uu I_{k_1}\dots \uu
I_{k_n}$ replaced by $Z_{k_1}\dots Z_{k_n}$. Choosing the initial and
final wavepackets $f_j$, $g_j$ normalized with respect to the scalar
product $\langle\cdot|\cdot\rangle_Z$ from Theorem 1 we find that the
modificaltion of the amplitude is of precisely the type we have
formulated in the Conjecture: Coupling constants in momentum space
are regularized according to $g_k\to g_kZ_k^{\frac{1}{2}}$ and the
ultraviolet infinities are regularized by square summability of
$Z_k$.

\subsection{The physical meaning of the parameters occuring in the
multi-oscillator non-CCR representation}

The Hamiltonian (\ref{ipH}) can be also
written as
\be
\uu H(t)&=&
i\frac{\hbar\alpha}{2}
\sum_{\bbox k}
\sqrt{\frac{\hbar}{2\omega_k}}\Big(
R_+\oplus_{\frac{1}{\sqrt{N}}}a{_{\bbox k}}e^{-i\Delta_{\bbox k} t}
-
R_-\oplus_{\frac{1}{\sqrt{N}}}a{_{\bbox k}}^{\dag}e^{i\Delta_{\bbox k} t}
\Big).
\ee
The choice of $\alpha_N=\frac{1}{\sqrt{N}}$ was motivated by purely formal
reasons. With this choice: (1) the RHS of field commutators satisfied
the resolution of unity, (2) the statistics of coherent-state
excitations was Poissonian in the thermodynamic limit, (3) the
formfactors $Z(\bbox k)$ were regularizing the interaction term in
the thermodynamic limit, (4) non-canonical coherent-state averages at the
multi-oscillator and single-oscillator levels were identical, (5)
normalization of single-photon and multi-photon states could be given
in terms of the same scalar product
$\langle \cdot|\cdot\rangle_Z$ in the thermodynamic limit.

The restriction $\uu H(t)_N$ of $\uu H(t)$ to an
$N$-oscillator subspace reads
\be
\uu H(t)_N&=&
\sum_{j=1}^N
i\frac{1}{\sqrt{N}}\frac{\hbar\alpha}{2}
\sum_{\bbox k}
\sqrt{\frac{\hbar}{2\omega_k}}\Big(
R_+
a_{\bbox k}^{(j)}
e^{-i\Delta_{\bbox k} t}
-
R_-
a_{\bbox k}^{(j)\dag}e^{i\Delta_{\bbox k} t}
\Big)
\ee
where $a_{\bbox k}^{(j)}=I\otimes\dots a_{\bbox
k} \dots \otimes I$ with the single-oscillator annihilation operator
at the $j$th place.

It is clear that $\uu H(t)_N$ represents the interaction Hamiltonian
of a single 2-level system interacting with $N$ indefinite-frequency
harmonic oscillators. The factor $\frac{1}{\sqrt{N}}$ means that the
strength of the coupling decreases with growing $N$, a property whose
physical interpretation is well known and occurs, for example,
in the Hepp-Lieb treatment of the Dicke model \cite{Dicke,HL}.

The Dicke model is in a sense dual to the one we consider. It
represents a single harmonic oscillator interacting with $N$ 2-level
systems via the interaction-picture Hamiltonian
\be
H(t)_{N\rm Dicke}
&=&
\sum_{j=1}^N
i\frac{1}{\sqrt{N}}\frac{\hbar\alpha}{2}
\sqrt{\frac{\hbar}{2\Omega}}\Big(
R_+^{(j)}
a_{\Omega}
e^{-i\Delta_{\Omega} t}
-
R_-^{(j)}
a_{\Omega}^{\dag}e^{i\Delta_{\Omega} t}
\Big).\label{Dicke1}
\ee
Here $R_\pm^{(j)}=I\otimes\dots R_\pm \dots \otimes I$,
and $[a_\Omega,a_\Omega^{\dag}]=I$.

Now, assuming the particular representation of
$[a_\Omega,a_\Omega^{\dag}]=I$ given by
$\Omega=\sum_k\omega_k|\bbox k\rangle\langle \bbox k|\otimes 1$,
$I=1\otimes 1$,
$[\hat a_{k},\hat a_{l}^{\dag}]=\delta_{kl}1$,
$a_k=|\bbox k\rangle\langle \bbox k|\otimes\hat a_k$,
$a_\Omega=\sum_k a_k$,
we can rewrite
(\ref{Dicke1}) as
\be
H(t)_{N\rm Dicke}
&=&
\sum_{j=1}^N
i\frac{1}{\sqrt{N}}\frac{\hbar\alpha}{2}
\sum_{\bbox k}
\sqrt{\frac{\hbar}{2\omega_k}}\Big(
R_+^{(j)}
a_{\bbox k}
e^{-i\Delta_{\bbox k} t}
-
R_-^{(j)}
a_{\bbox k}^{\dag}e^{i\Delta_{\bbox k} t}
\Big).\label{Dicke2}
\ee
Although the Hamiltonian (\ref{Dicke1}) is formally a {\it
single-mode\/} one, the form
(\ref{Dicke2}) resembles the non-canonical interaction term discussed
in the previous subsection. What is important, the $N$-dependent
coupling constant can be derived on physical grounds if one assumes a
constant density $N/V$ of (bosonic) atoms. To see this, one writes
the parameters of the Dicke-Hepp-Lieb Hamiltonian as
\be
\alpha(\hbar/2)^{3/2}
=
\omega_0d \sqrt{2\pi \hbar\rho}
\ee
where $\rho=N/V$ is the density of atoms, $d$ the dipole moment, and
$\omega_0$ the atomic frequency. The termodynamic limit of the
Hepp-Lieb approach is performed under the assumption $\alpha={\rm
const}$ which is equivalent to the constant density $\rho$.

This is precisely what happens in our approach to the thermodynamic
limit.
It follows that the choice of the non-CCR algebra whose RHS satisfies the
resolution of unity is physically equivalent to the requirement that
the electromagnetic field consists of indefinite-frequency
oscillators of constant density $N/V$. The thermodynamic limit is then
equivalent to the infinite volume limit $V\to\infty$.

Finally, let us note that taking the direct sum over $N$ we find
\be
\uu H(t)_{\rm Dicke}
&=&
\oplus_{N=1}^\infty H(t)_{N\rm Dicke}\nonumber\\
&=&
i\frac{\hbar\alpha}{2}
\sum_{\bbox k}
\sqrt{\frac{\hbar}{2\omega_k}}\Big(
\uu R_+
a_{\bbox k}
e^{-i\Delta_{\bbox k} t}
-
\uu R_-
a_{\bbox k}^{\dag}e^{i\Delta_{\bbox k} t}
\Big)\label{Dicke3}
\ee
where $\uu R_\pm =\oplus_{\frac{1}{\sqrt{N}}}R_\pm$, i.e. a structure
analogous to the interaction Hamiltonian in non-canonical quantum
optics.

\subsection{Structure of the space of states}

The Conjecture we have formulated above implies that probabilities have to
be computed as if the vacuum was unique. However, it is essential for
the non-canonical construction to have an infinite dimensional space
of different vacua. Otherwise it would be impossible to associate the
spectrum of frequencies with a single harmonic oscillator. There is
no contradiction if we treat the Fock space genereated from a vacuum
state $|\uu O\rangle$ as a fiber over $|\uu O\rangle$ in a vector
bundle with Fock fibers and the base space consisting of vacua.
Vacuum fluctuations represented by the zero-energy part of the
Hamiltonian generate a motion (a flow) in the base space and thus
play a role of bundle connection. The ``vacuum picture" corresponds
to a choice of connection.

\section{Conclusions and further perspectives}

Fields with the property of having at the RHS of the commutator an
element from a nontrivial center of the algebra are usually
termed the generalized free fields.
In the context of our formalism we find this term misleading
for two main reasons. First of all
the generalized free fields are ``free" whereas the examples we
discussed include scattering of radiation, spontaneous emission and
interaction picture perturbation theory. All these processes
involve interactions with charges. Secondly, we arrived at the
particular non-CCR representation on the basis of a concrete physical
model of the indefinite frequency oscillator. The mathematical
structure of the model does not follow the typical limitations
imposed on generalized free fields.
For example, the vacuum states are Poincar\'e non-invariant.
The same concerns the RHS of
the non-CCR algebra which is only Poincar\'e covariant
and satisfies the resolution of unity.
The issues of locality have to be formulated in terms of certain operator
generalizations of Green and Jordan-Pauli functions which lead to
very special locality properties and eliminate problems with
multiplication of field operators at the same point in configuration space.
The notion of the thermodynamic limit is typical of the
representation we use and does not seem to have been investigated in
the context of axiomatic quantum fields.

The construction we are developing has a slightly different logic
than the usual axiomatic quantum field theory. It seems that what we
are doing is somehow in-between the generalized free fields, nonlocal
quantum field theory, and noncommutative geometry.

An issue which has not been addressed so far is how to quantize
fermions. Some results on the Dirac equation
are known already and will be presented in a
separate paper.

Another problem is to embed the concrete non-canonical
quantization procedure we have proposed into a more abstract scheme
of quantizations in a $C^*$-algebraic setting. The fact that the
right-hand-side of commutation relations is not an identity but
rather an operator belonging to the center of the algebra suggests
directions for generalizations. It seems there is a link with the
work of Streater on non-abelian cocycles \cite{Streater}.
An appropriate version
of a coherent-state quantization based on the formalism of Naudts and
Kuna \cite{NK} is in preparation.

\acknowledgments
The work was done mainly during my stays in Antwerp and Clausthal
with NATO and Alexander-von-Humboldt fellowships. I
am indebted to Prof. H.-D. Doebner, W. L\"ucke, J. Naudts and M. Kuna for many
interesting discussions.


\begin{references}
\bibitem{PT}In this paper when we speak of the Poincar\'e group we
mean the semidirect product of 4-translations and $SL(2,C)$, i.e. the
universal covering space of the Poincar\'e group.
\end{references}
\end{document}